# Analysis of Genomic and Transcriptomic Variations as Prognostic Signature for Lung Adenocarcinoma

Talip Zengin, Tuğba Önal-Süzek✉

## Abstract

### Background

Lung cancer is the leading cause of largest number of deaths worldwide and lung adenocarcinoma is the most common form of lung cancer. In order to understand the molecular basis of lung adenocarcinoma, integrative analyses have been performed by using genomics, transcriptomics, epigenomics and clinical data. Besides, molecular prognostic signatures have been generated for lung adenocarcinoma by using gene expression levels in tumor samples. However, we need signatures including different types of molecular data, even cohort or patient-based biomarkers which are candidate of molecular targeting.

### Results

We built an R pipeline to carry out an integrated meta-analysis of the genomic alterations including single-nucleotide variations and the copy number variations, transcriptomics variations through RNA-seq and clinical data of patients with lung adenocarcinoma in The Cancer Genome Atlas project. We integrated significant genes including single-nucleotide variations or the copy number variations, differentially expressed genes and those in active subnetworks to construct prognosis signature. Cox proportional hazards model with Lasso penalty and LOOCV was used to identify best gene signature among different gene categories.

We determined 12-gene signature (BCHE, CCNA1, CYP24A1, DEPTOR, MASP2, MGLL, MYO1A, PODXL2, RAPGEF3, SGK2, TNNI2, ZBTB16) for prognostic risk prediction based on overall survival time of the patients with lung adenocarcinoma. The patients in both training and test data were clustered into high-risk and low-risk groups by using risk scores of the patients calculated based on selected gene signature. Overall survival probability of these risk groups was highly significantly different for both training and test datasets.

## Conclusions

These 12-gene signature could predict the prognostic risk of the patients with lung adenocarcinoma in TCGA and they are potential predictors for the survival-based risk clustering of the patients with lung adenocarcinoma. These genes can be used to cluster patients based on molecular nature and the best candidates of drugs for the patient clusters can be proposed. These genes also have high potential for targeted cancer therapy of patients with lung adenocarcinoma.

## Keywords

TCGA, Lung cancer, Lung adenocarcinoma, Differential Expression, SNV, CNV, Active Subnetwork, Cox Proportional Hazards Regression, Signature, Survival

## Background

Lung cancer is the most common cancer and responsible for largest number of deaths worldwide with 1.8 million deaths, 18.4% of the total (IARC, 2018). Lung cancer is categorized into two main categories: non-small cell lung cancer (NSCLC) which occurs in 85% of patients and small cell lung cancer (SCLC) in 15% of cases. NSCLC is grouped into 3 histological sub-types: lung adenocarcinoma (LUAD) which is most common form of lung cancer, lung squamous cell carcinoma (LUSC) and large cell carcinoma (Travis, 2011).

Integration of different types of molecular data has been used to characterize molecular basis of lung cancer and to determine clinical status of patients. Shi et al. analyzed 101 LUAD samples by using data from different levels -DNA mutations, gene expression profile, copy number variations and DNA methylation- in order to identify the relation between genomic status and clinical status. They determined deleterious mutations at ZKSCAN1 and POU4F2 genes which are two novel candidate driver genes (Shi, 2016). Furthermore, recent studies have been performed to generate new methods to analyze integrative cancer data. Berger et al. proposed a new method called expression-based variant-impact phenotyping (eVIP) using differentially expressed genes (DEGs) to distinguish driver mutations from passenger mutations. They characterized 194 somatic mutations related with primary LUAD and claimed that 69% of mutations were mutations. They found the driver mutations in LUAD are EGFR (p.S645C), ERBB2 (p.S418T), ARAF (p.S214C) and ARAF (p.S214F) (Berger, 2016). TCGA network analyzed 230 LUAD samples using mRNA, microRNA and DNA sequencing integrated with copy number, methylation and proteomic data (The Cancer Genome Atlas Research Network, 2014) and reported the samples with high rates of somatic mutation. Eighteen genes with high mutation load were reported such as RIT1 activating mutations and MGA loss-of-function mutations. They also identified aberrations in NF1, MET, ERBB2 and RIT1 occurred in 13% of cases and MAPK and PI(3)K pathway activity (The Cancer Genome Atlas Research Network, 2014). Deng Z. et al., presented genomic alterations in LUAD samples from TCGA and found the significantly aberrant CNV segments which are associated with the immune system and 63 mutated genes associated with lung cancer signaling related to cancer progression. They identified important mutations of the PI3K protein family members include PIK3C2B, PIK3CA, PIK3R1 (Deng, 2017).

Recently, studies have been performed to generate gene signatures predicting prognosis risk of patients with lung adenocarcinoma. In 2016, Krzystanek et al. identified 7-gene signature by using microarray data of early stage lung adenocarcinoma from GEO datasets. The genes (ADAM10, DLGAP5, RAD51AP1, FGFR10P, NCGAP, KIF15, ASPM) which have high hazards ratios showed significant results at cox regression analysis and Kaplan-Meier survival plots (Krzystanek, 2016). Shukla et.al. identified 96 genes including five long noncoding RNAs (lncRNAs) among training data which had prognostic association at test data, by using lung adenocarcinoma RNA-seq and clinical data from TCGA (Shukla, 2017). Shi et.al. studied on long noncoding RNAs (lncRNAs) expression signature model to predict stage I lung adenocarcinoma from TCGA and determined 31-lncRNA signature to predict overall survival in patients with LUAD (Shi, 2018). Zhao et. al. used gene expression profiles from TCGA and identified 20 genes which were significantly associated with the overall survival (OS). When they combined with GEO data set, they obtained four genes, FUT4, SLC25A42, IGFBP1, and KLHDC8B as common (Zhao, 2018). Li et. al. performed RNA-sequencing on LUAD tumor samples and normal tissue samples. They construct protein–protein interaction network by using DEGs which were intersection of GEO datasets and identified hub genes. Then, they test these genes on patient cohorts and TCGA data. They identified eight genes (DLGAP5, KIF11, RAD51AP1, CCNB1, AURKA, CDC6, OIP5 and NCAPG) which were closely related to survival in LUAD (Li, 2018). He et. al. studied on previous GEO datasets and TCGA data and they identified 8-gene prognostic signature (CDCP1, HMMR, TPX2, CIRBP, HLF, KBTBD7, SEC24B-AS1, and SH2B1) by using the step-wise multivariate Cox analysis. These genes were good predictors of survival between high-risk and low-risk groups of patients with early-stage NSCLC (He, 2019). The studies above determined different gene signatures for prognosis risk prediction by using different methods and presented different genes. Although, mostly gene expression data has been used for this purpose, we integrated SNVs, CNVs, DEGs and active subnetwork DEGs to generate gene signature for risk model by using LUAD data from The Cancer Genome Atlas (TCGA) database which provides simple nucleotide variation, gene expression, miRNA expression, DNA methylation, copy number variation and reverse phase protein array, clinical and biospecimen data from more than 10,000 cancer patients with 39 cancer types (Chang, 2013).

In this study, we built an R pipeline (Figure 19) to perform an integrative analysis including SNVs and CNVs, differentially expressed genes and clinical data of patients with lung adenocarcinoma in TCGA. We generated different data categories by using significant SNVs, CNVs, DEGs and active subnetwork DEGs. Multivariate Cox proportional hazards model with Lasso penalty and LOOCV was used to identify best gene signature among different gene categories. We generated 12-gene signature (BCHE, CCNA1, CYP24A1, DEPTOR, MASP2, MGLL, MYO1A, PODXL2, RAPGEF3, SGK2, TNNI2, ZBTB16) for prognostic risk prediction based on overall survival time of the patients with lung adenocarcinoma. When we clustered patients into high-risk and low-risk groups, the survival analysis showed highly significant results for both training and test datasets.

## Results

### Identification of Significant Simple Nucleotide Variations

Mutation data of LUAD patients as *maf* file generated by *mutect* pipeline was downloaded by *TCGAbiolinks* package and *maftools* package was used to subset original *maf* file by tumor sample barcodes of 55 LUAD patients and 510 LUAD patients. Then, significant mutations for both 55 and 510 LUAD patients were determined separately with their roles as tumor suppressor or oncogene by *SomInaClust* R package. In order to determine important genes including significant mutation clusters, we used SomInaClust R package. EGFR, KRAS, TP53, STK11, RB1 and MGA genes were determined as candidate driver genes in tumor samples of 55 LUAD patients (Figure 1). EGFR and KRAS genes were classified as oncogene and STK11, RB1 and MGA genes were classified as tumor suppressor. Although TP53 gene has both OG score and TSG score, TP53 was classified as tumor suppressor in Table 1 depending on reference information of cancer gene census. EGFR, KRAS, TP53, STK11 and RB1 have highly significant estimation. While EGFR and TP53 have higher number of mutations, KRAS, STK11, RB1 and MGA have lower number of mutations. While EGFR, KRAS, TP53, STK11, RB1 are well known cancer related genes, MGA gene is not in cancer gene census.

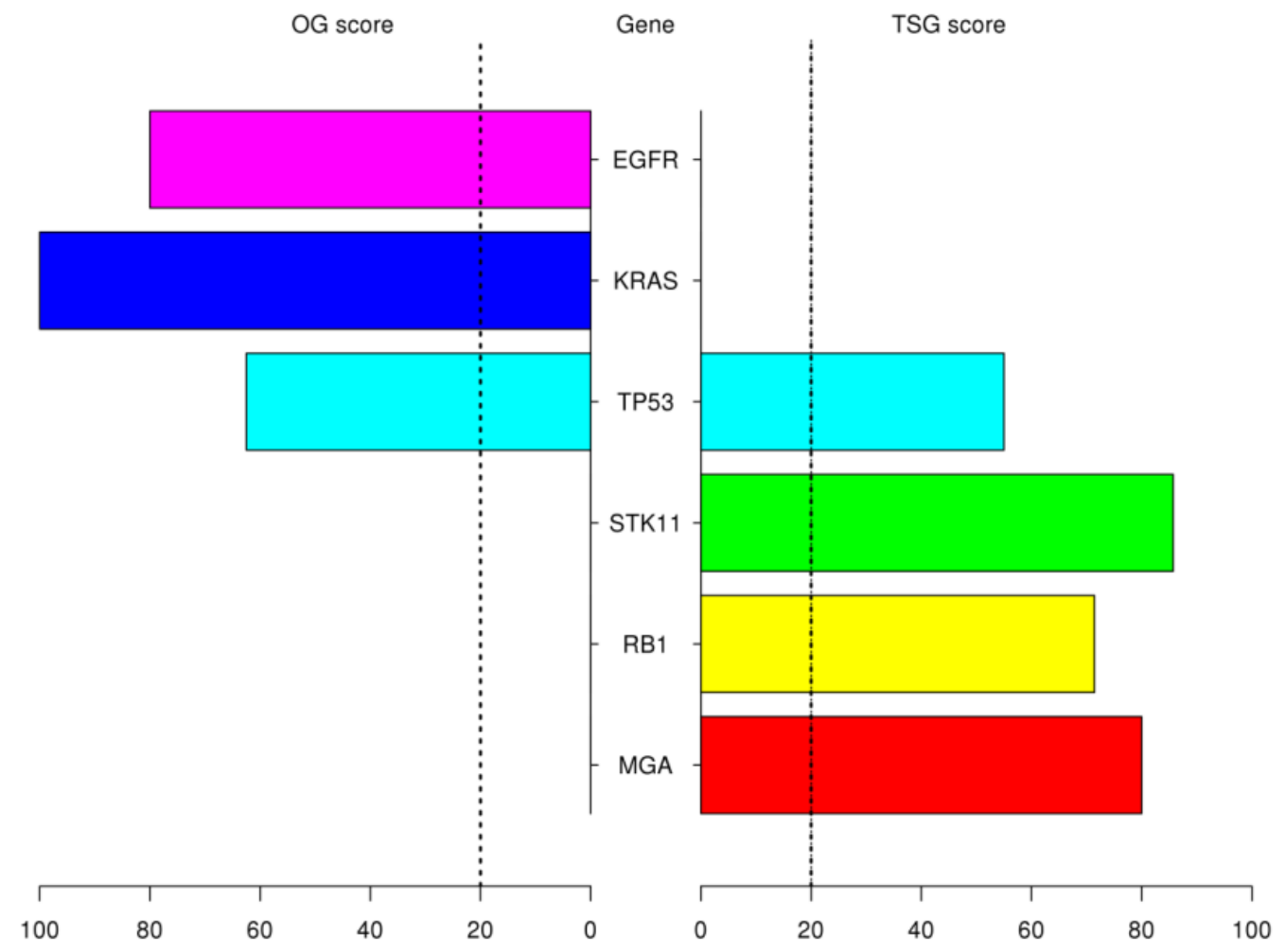

**Figure 1. Pyramid plot of important mutated genes classified as oncogene (OG) or tumor suppressor gene (TSG) in tumor samples of 55 patients with LUAD**

**Table 1. Significant mutated genes in 55 tumor samples**

| Gene | # Mutations | Q value | OG Score | TSG Score | Classification | CGC* |
|---|---|---|---|---|---|---|
| EGFR | 11 | 1.57e-12 | 80 | 0 | OG | Dom |
| KRAS | 8 | 1.57e-12 | 100 | 0 | OG | Dom |
| TP53 | 20 | 4.8e-07 | 62.5 | 55 | TSG | Rec |
| STK11 | 7 | 0.000106 | 0 | 85.7 | TSG | Rec |
| RB1 | 7 | 0.0049 | 0 | 71.4 | TSG | Rec |
| MGA | 6 | 0.0217 | 0 | 80 | TSG | NA |

* Cancer gene census (Dom: Dominant, Rec: Recessive)

Eighty-two genes as candidate driver genes in tumor samples of 510 LUAD patients (Table 2), including KRAS, TP53, EGFR, STK11, MGA and RB1 which were determined also in tumor samples of 55 LUAD patients (Figure 2). These genes include very well-known cancer related oncogenes such as BRAF, ERBB2, AKT1 and PIK3CA with the genes which are not listed in cancer gene census list of COSMIC database (Table 2).

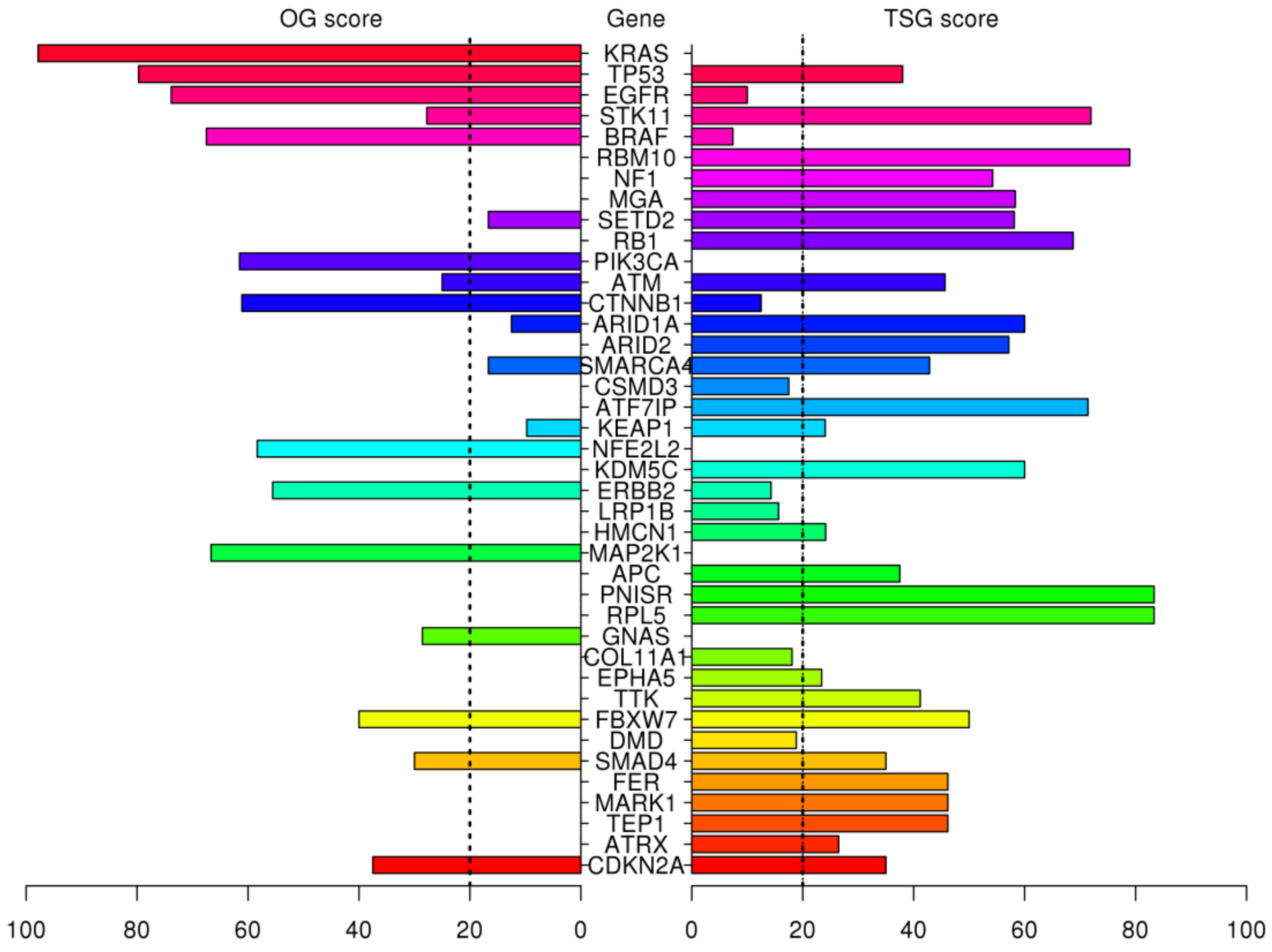

**Figure 2. Pyramid plot of top 40 important mutated genes classified as oncogene (OG) or tumor suppressor gene (TSG) in tumor samples of 510 patients with LUAD**

**Table 2. Significant mutated genes in tumor samples of 510 patients with LUAD**

| Gene | # Mutations | qDG | OG Score | TSG Score | Classification | CGC* |
|---|---|---|---|---|---|---|
| KRAS | 143 | 1.97e-250 | 97.8 | 0 | OG | Dom |
| TP53 | 253 | 2.52e-135 | 79.7 | 38 | TSG | Rec |
| EGFR | 73 | 8.97e-84 | 73.8 | 10 | OG | Dom |
| STK11 | 83 | 4.6e-61 | 27.8 | 72 | TSG | Rec |
| BRAF | 44 | 8.07e-51 | 67.5 | 7.4 | OG | Dom |
| RBM10 | 39 | 9.06e-31 | 0 | 78.9 | TSG | NA |
| NF1 | 63 | 5.37e-25 | 0 | 54.2 | TSG | Rec |
| MGA | 52 | 6.46e-23 | 0 | 58.3 | TSG | NA |
| SETD2 | 44 | 1.34e-20 | 16.7 | 58.1 | TSG | Rec |
| RB1 | 32 | 4.99e-20 | 0 | 68.8 | TSG | Rec |
| PIK3CA | 27 | 1.36e-19 | 61.5 | 0 | OG | Dom |
| ATM | 48 | 5.18e-18 | 25 | 45.7 | TSG | Rec |
| CTNNB1 | 21 | 3.32e-15 | 61.1 | 12.5 | OG | Dom |
| ARID1A | 30 | 1.76e-14 | 12.5 | 60 | TSG | Rec |
| ARID2 | 29 | 2.83e-12 | 0 | 57.1 | TSG | Rec |
| SMARCA4 | 48 | 2.23e-11 | 16.7 | 42.9 | TSG | Rec |
| CSMD3 | 324 | 6.25e-10 | 0 | 17.5 | NA | NA |
| ATF7IP | 17 | 1.84e-08 | 0 | 71.4 | TSG | NA |
| KEAP1 | 90 | 1.91e-08 | 9.8 | 24.1 | TSG | NA |
| NFE2L2 | 14 | 2.83e-07 | 58.3 | 0 | OG | Dom |
| KDM5C | 16 | 1.76e-06 | 0 | 60 | TSG | Rec |
| ERBB2 | 13 | 6.94e-06 | 55.6 | 14.3 | OG | Dom |
| LRP1B | 267 | 6.04e-05 | 0 | 15.6 | NA | Rec |
| HMCN1 | 97 | 8.93e-05 | 0 | 24.1 | TSG | NA |
| MAP2K1 | 9 | 0.000263 | 66.7 | 0 | OG | Dom |
| APC | 24 | 0.000272 | 0 | 37.5 | TSG | Rec |
| PNISR | 6 | 0.000626 | 0 | 83.3 | TSG | NA |
| RPL5 | 7 | 0.000626 | 0 | 83.3 | TSG | Dom |
| GNAS | 19 | 0.000962 | 28.6 | 0 | OG | Dom |
| COL11A1 | 129 | 0.00139 | 0 | 18.1 | NA | NA |
| EPHA5 | 66 | 0.00221 | 0 | 23.4 | TSG | NA |
| TTK | 18 | 0.00221 | 0 | 41.2 | TSG | NA |
| FBXW7 | 12 | 0.0028 | 40 | 50 | TSG | Rec |
| DMD | 99 | 0.00349 | 0 | 18.8 | NA | NA |
| SMAD4 | 20 | 0.00379 | 30 | 35 | TSG | Rec |
| FER | 16 | 0.0043 | 0 | 46.2 | TSG | NA |
| MARK1 | 21 | 0.0043 | 0 | 46.2 | TSG | NA |
| TEP1 | 29 | 0.0043 | 0 | 46.2 | TSG | NA |
| ATRX | 35 | 0.00463 | 0 | 26.5 | TSG | Rec |
| CDKN2A | 21 | 0.00585 | 37.5 | 35 | TSG | Rec |
| MYO9A | 19 | 0.00615 | 0 | 42.9 | TSG | NA |
| ZNF800 | 17 | 0.00615 | 0 | 42.9 | TSG | NA |
| CMTR2 | 26 | 0.00674 | 0 | 55.6 | TSG | NA |
| RASA1 | 9 | 0.00674 | 0 | 55.6 | TSG | NA |
| CDKN1B | 5 | 0.00674 | 0 | 80 | TSG | Rec |
| DHX15 | 7 | 0.00674 | 0 | 80 | TSG | NA |
| IQGAP2 | 28 | 0.00816 | 0 | 40 | TSG | NA |
| LTN1 | 19 | 0.00816 | 0 | 40 | TSG | NA |
| SMARCA1 | 19 | 0.00816 | 0 | 40 | TSG | NA |
| SPTA1 | 164 | 0.00971 | 0 | 17.6 | NA | NA |
| FHOD3 | 31 | 0.0122 | 0 | 30.4 | TSG | NA |
| CPVL | 8 | 0.0161 | 0 | 66.7 | TSG | NA |
| MAP3K12 | 8 | 0.0161 | 0 | 66.7 | TSG | NA |
| TOP2B | 9 | 0.0161 | 0 | 66.7 | TSG | NA |
| ROCK1 | 21 | 0.0163 | 0 | 35.3 | TSG | NA |
| PBRM1 | 12 | 0.0172 | 0 | 45.5 | TSG | Rec |
| AKAP6 | 40 | 0.0195 | 0 | 28 | TSG | NA |
| SENP1 | 3 | 0.0241 | 0 | 100 | TSG | NA |
| SP1 | 4 | 0.0241 | 0 | 100 | TSG | NA |
| WISP3 | 4 | 0.0241 | 0 | 100 | TSG | NA |
| RAD50 | 13 | 0.0243 | 20 | 41.7 | TSG | NA |
| COL28A1 | 19 | 0.0243 | 0 | 41.7 | TSG | NA |
| SCAF8 | 18 | 0.0243 | 0 | 41.7 | TSG | NA |
| STK31 | 19 | 0.0243 | 0 | 41.7 | TSG | NA |
| IDH1 | 6 | 0.0248 | 40 | 25 | TSG | Dom |
| USH2A | 240 | 0.0263 | 0 | 13.2 | NA | NA |
| YLPM1 | 23 | 0.0269 | 0 | 31.6 | TSG | NA |
| IQUB | 12 | 0.0272 | 0 | 57.1 | TSG | NA |
| MARK2 | 10 | 0.0272 | 0 | 57.1 | TSG | NA |
| NAA15 | 8 | 0.0272 | 0 | 57.1 | TSG | NA |
| CDH10 | 99 | 0.028 | 0 | 16.4 | NA | NA |
| AKT1 | 3 | 0.0296 | 66.7 | 0 | OG | Dom |
| RAF1 | 7 | 0.031 | 66.7 | 0 | OG | Dom |
| VPS13C | 39 | 0.0332 | 0 | 25 | TSG | NA |
| ZBBX | 28 | 0.0333 | 0 | 30 | TSG | NA |
| DST | 67 | 0.0333 | 0 | 19.1 | NA | NA |
| KMT2C | 52 | 0.0388 | 0 | 18.8 | NA | Rec |
| DGKB | 38 | 0.0431 | 0 | 28.6 | TSG | NA |
| MAP2K4 | 8 | 0.045 | 33.3 | 50 | TSG | Rec |
| FBN2 | 93 | 0.045 | 0 | 20.5 | TSG | NA |
| B2M | 8 | 0.045 | 0 | 50 | TSG | Rec |
| BAP1 | 8 | 0.045 | 0 | 50 | TSG | Rec |

* Cancer gene census (Dom: Dominant, Rec: Recessive)

## Identification of Significant Copy Number Variations

CNVs (Copy Number Variations) are important aberrations which results alterations in gene expression in tumorigenesis and tumor growth. In order to determine significant CNVs among tumor samples of 55 and 510 LUAD patients, *gaia* R package was used. Significant recurrent CNVs in tumor samples of 55 LUAD patients, over the q-value thresholds (0.01), are mostly observed on Chromosome 1, 8, 9, and 17. Chromosome 1 has the highest number of amplifications followed by Chromosome 8. Chromosome 9 has the highest number of deletions followed by Chromosome 17 as seen in Figure 3. Chromosome 1 has the highest number of gene aberration with 2006 amplified or deleted genes followed by Chromosome 8 with 1029 aberrant genes and Chromosome 19 with 785 aberrant genes. Top ten significant amplified and deleted genes which are all from chromosome 1 are listed in Table 3.

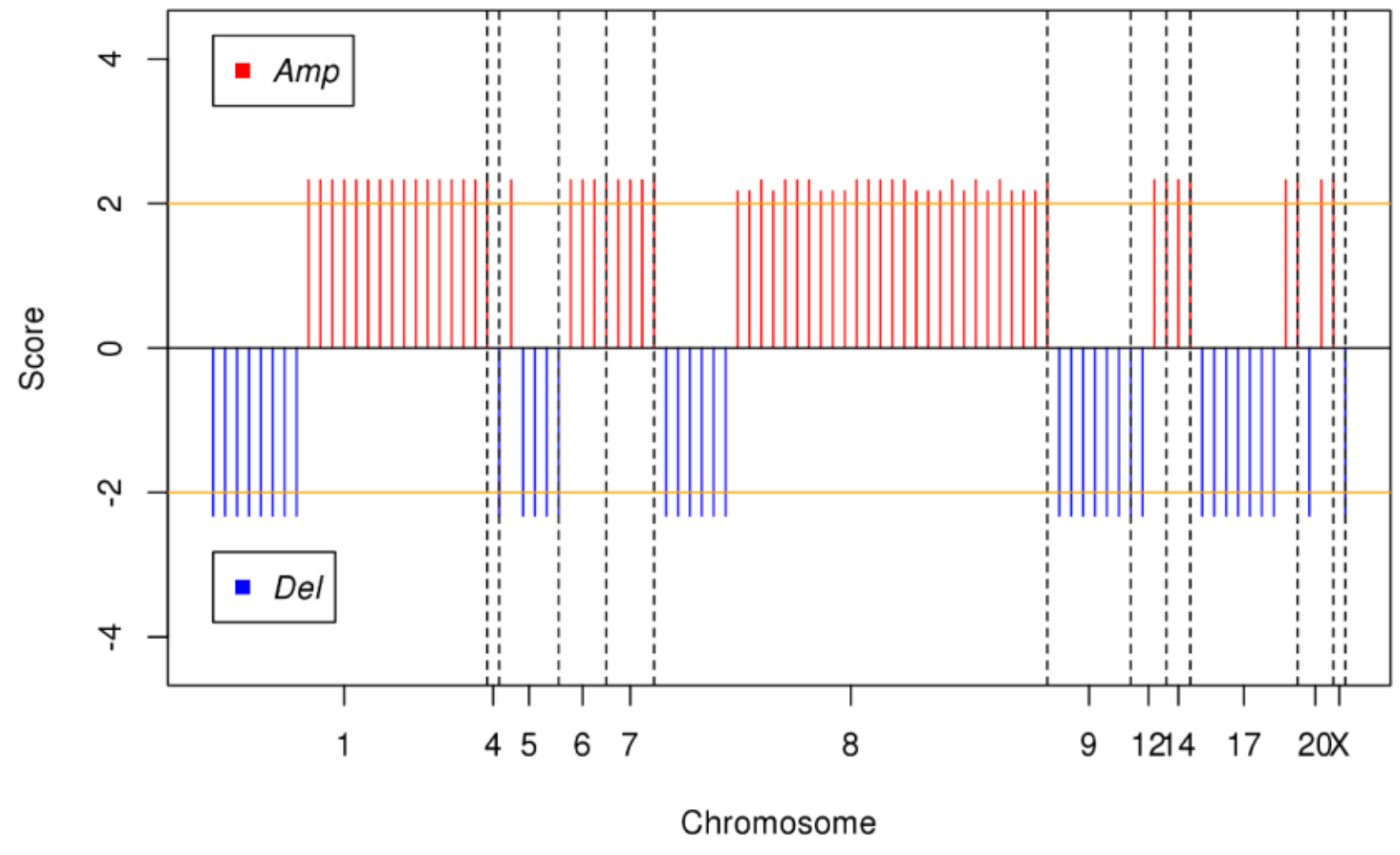

**Figure 3. Significant CNVs on all chromosomes in tumor samples of 55 patients with LUAD**

**Table 3. Top ten significant deleted and amplified genes in tumor samples of 55 patients with LUAD**

| Gene Symbol | Aberration | q-value | Aberrant Region | Gene Region |
|---|---|---|---|---|
| RN7SKP285 | Del | 0.00474651 | 1:103501576-107318961 | 1:103523562-103523879 |
| RNPC3 | Del | 0.00474651 | 1:103501576-107318961 | 1:103525691-103555239 |
| AMY2B | Del | 0.00474651 | 1:103501576-107318961 | 1:103553815-103579534 |
| ACTG1P4 | Del | 0.00474651 | 1:103501576-107318961 | 1:103569553-103570674 |
| AMY2A | Del | 0.00474651 | 1:103501576-107318961 | 1:103616811-103625780 |
| AMY1A | Del | 0.00474651 | 1:103501576-107318961 | 1:103655290-103664554 |
| AC105272.1 | Del | 0.00474651 | 1:103501576-107318961 | 1:103668071-103668268 |
| AMY1B | Del | 0.00474651 | 1:103501576-107318961 | 1:103687415-103696680 |
| AMYP1 | Del | 0.00474651 | 1:103501576-107318961 | 1:103713723-103719871 |
| AMY1C | Del | 0.00474651 | 1:103501576-107318961 | 1:103750406-103758690 |
| PLEKHO1 | Amp | 0.00474651 | 1:150131878-150768299 | 1:150149183-150164720 |
| AC242988.2 | Amp | 0.00474651 | 1:150131878-150768299 | 1:150173049-150181429 |
| RN7SL480P | Amp | 0.00474651 | 1:150131878-150768299 | 1:150211632-150211925 |
| ANP32E | Amp | 0.00474651 | 1:150131878-150768299 | 1:150218417-150236156 |
| RNU2-17P | Amp | 0.00474651 | 1:150131878-150768299 | 1:150236967-150237156 |
| AC242988.1 | Amp | 0.00474651 | 1:150131878-150768299 | 1:150255095-150257286 |
| CA14 | Amp | 0.00474651 | 1:150131878-150768299 | 1:150257251-150265078 |
| APH1A | Amp | 0.00474651 | 1:150131878-150768299 | 1:150265399-150269580 |
| C1orf54 | Amp | 0.00474651 | 1:150131878-150768299 | 1:150268200-150280916 |
| CIART | Amp | 0.00474651 | 1:150131878-150768299 | 1:150282543-150287093 |

Significant recurrent CNVs in tumor samples of 510 LUAD patients, over the q-value thresholds (0.01), are mostly observed on Chromosome 4, 9, 10, 11, 12, 13, 14, 16, 18 and 20. But Chromosome 11 has the highest number of aberrations followed by Chromosome 9, 16 and 18. Chromosome 4, 9, 10, 12 and 16 has mostly amplifications (Figure 4). The pattern of CNVs in tumor samples of 510 patients has a marked difference from the CNV pattern in tumor samples of 55 patients (Figure 3). Chromosome 1 has the highest number of gene aberration with 3124 amplified or deleted genes followed by Chromosome 6 with 2911 aberrant genes and Chromosome 3 with 2149 aberrant genes. Top ten significant amplified and deleted genes which are all from chromosome 1 are showed in Table 4.

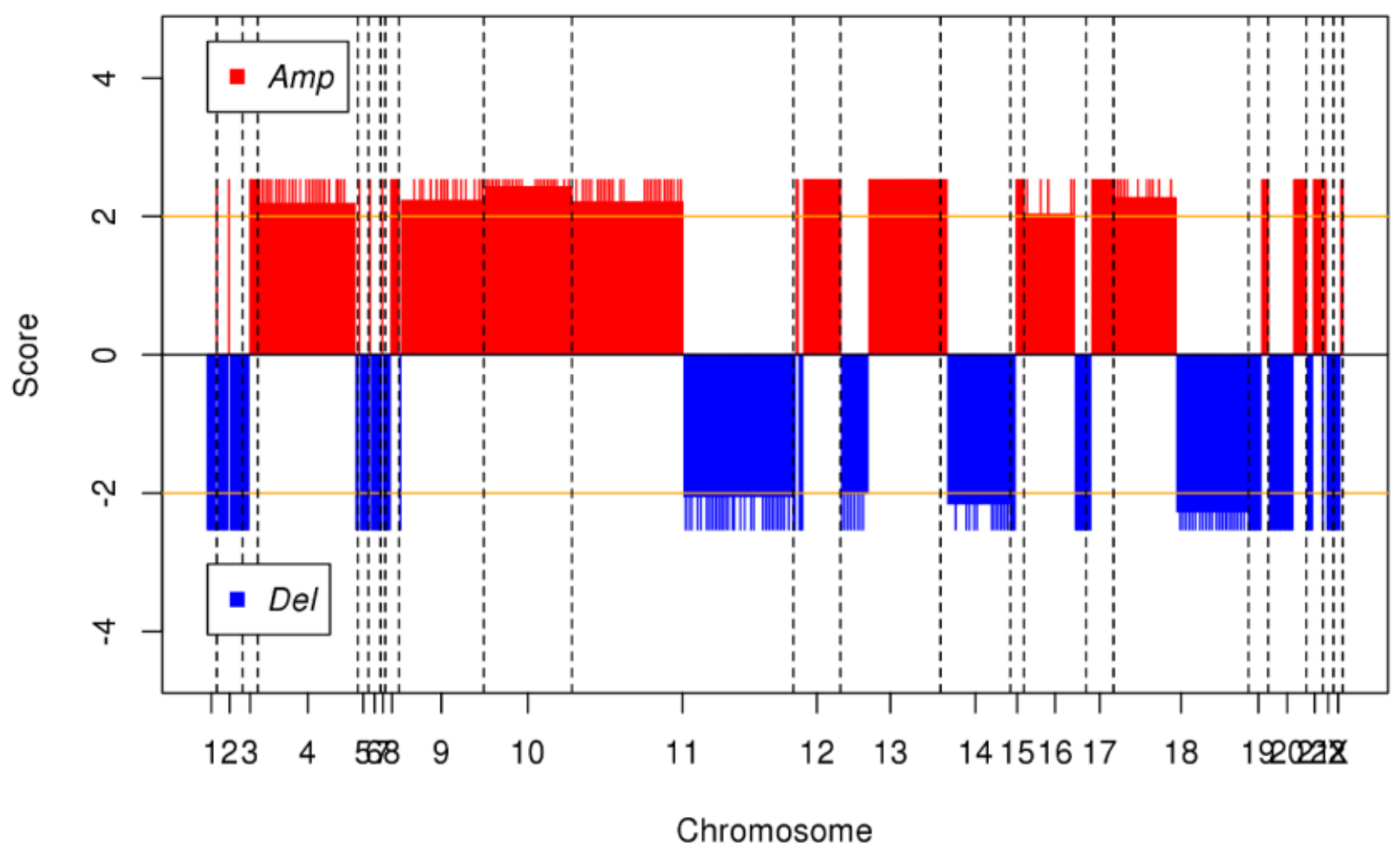

**Figure 4. Significant CNVs on all chromosomes in tumor samples of 510 patients with LUAD**

**Table 4. Top ten significant amplified and deleted genes in 510 LUAD patients**

| Gene Symbol | Aberration | q-value | Aberrant Region | Gene Region |
|---|---|---|---|---|
| AL359821.1 | Del | 0.0029609 | 1:71621685-71778398 | 1:71738173-71738354 |
| GDI2P2 | Del | 0.0029609 | 1:71928758-119984738 | 1:72274552-72275159 |
| AL513166.2 | Del | 0.0029609 | 1:71928758-119984738 | 1:72283170-72753772 |
| RPL31P12 | Del | 0.0029609 | 1:71928758-119984738 | 1:72301472-72301829 |
| AL583808.1 | Del | 0.0029609 | 1:71928758-119984738 | 1:72636547-72899240 |
| RNU6-1246P | Del | 0.0029609 | 1:71928758-119984738 | 1:72717663-72717769 |
| AL583808.2 | Del | 0.0029609 | 1:71928758-119984738 | 1:72765031-72791282 |
| AL583808.3 | Del | 0.0029609 | 1:71928758-119984738 | 1:72793104-72854475 |
| AL732618.1 | Del | 0.0029609 | 1:71928758-119984738 | 1:72979014-72979314 |
| KRT8P21 | Del | 0.0029609 | 1:71928758-119984738 | 1:73104792-73106282 |
| SF3B4 | Amp | 0.0029609 | 1:149907993-247650984 | 1:149923317-149927803 |
| MTMR11 | Amp | 0.0029609 | 1:149907993-247650984 | 1:149928651-149936879 |
| OTUD7B | Amp | 0.0029609 | 1:149907993-247650984 | 1:149937812-150010726 |
| AC244033.2 | Amp | 0.0029609 | 1:149907993-247650984 | 1:150045660-150067701 |
| AC244033.1 | Amp | 0.0029609 | 1:149907993-247650984 | 1:150053864-150055034 |
| VPS45 | Amp | 0.0029609 | 1:149907993-247650984 | 1:150067279-150145329 |
| PLEKHO1 | Amp | 0.0029609 | 1:149907993-247650984 | 1:150149183-150164720 |
| AC242988.2 | Amp | 0.0029609 | 1:149907993-247650984 | 1:150173049-150181429 |
| RN7SL480P | Amp | 0.0029609 | 1:149907993-247650984 | 1:150211632-150211925 |
| ANP32E | Amp | 0.0029609 | 1:149907993-247650984 | 1:150218417-150236156 |

## Differential Expression Analysis (DEA)

The Transcriptome Profiling data of LUAD patients in mRNA expression level (as unnormalized *HTSeq* raw counts), was downloaded by *TCGABiolinks* R package. Differentially expressed genes were determined with FDR adjusted p-values (q-values) in tumor samples (TP) of 55 patients with LUAD compared to normal samples (NT) of the same patients by *limma-voom* method using *limma* and *edgeR* R packages. The volcano plot in Figure 5, shows the differentially expressed genes (DEGs) as dots of which black ones represent the genes which have differential expression less than two-fold and not significant while red ones represent upregulated and green ones downregulated more than two-fold ($log_2$=1) significantly (q value < 0.01). As the result of this analysis, 3575 genes were dysregulated more than two-fold with 0.01 q-value significance.

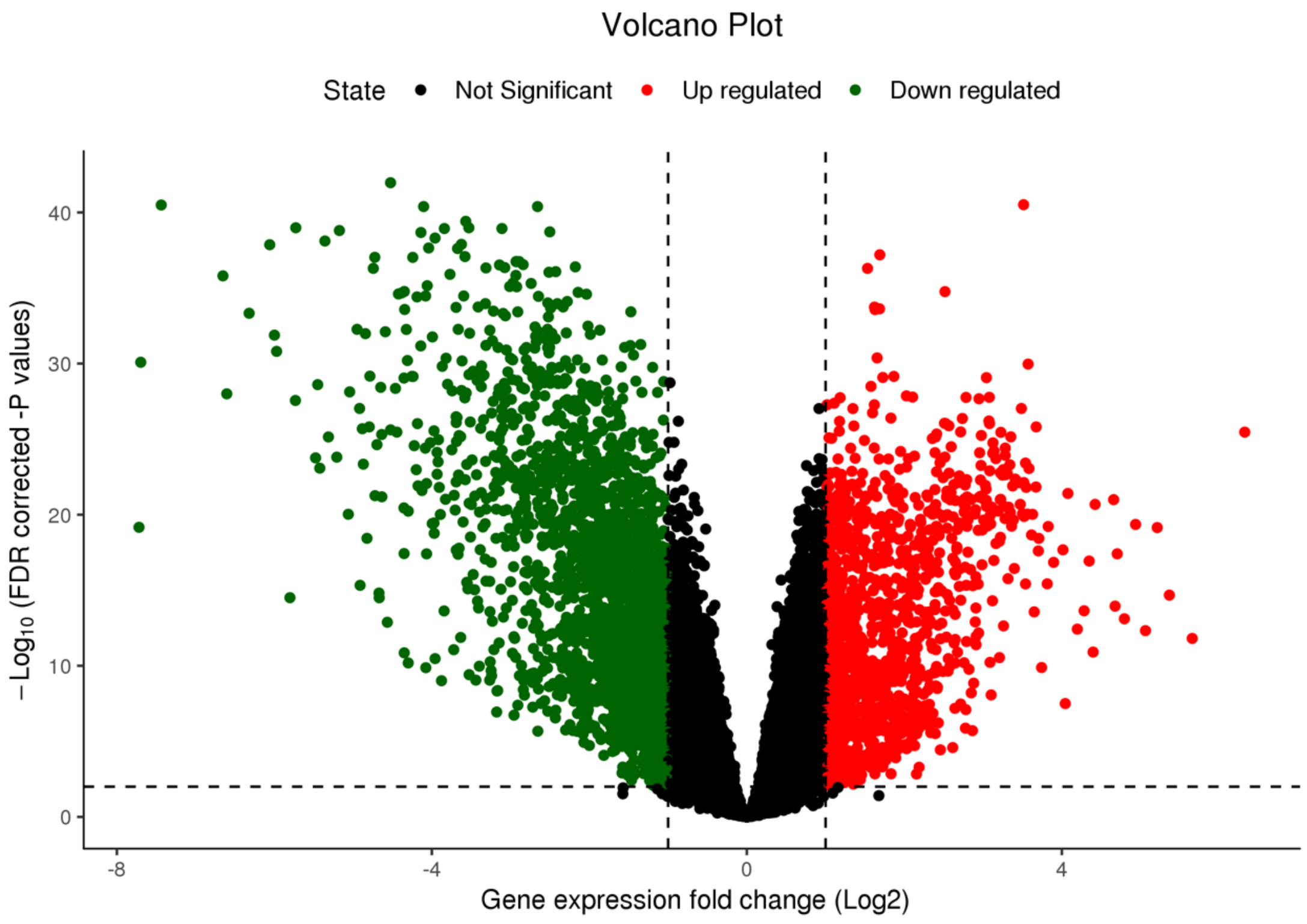

**Figure 5. Volcano plot of differentially expressed genes in tumor samples of 55 patients with LUAD.**

As the result of DEA, differentially expressed genes (DEGs) are determined with their log Fold Change (logFC), adjusted p-value (q-value), entrez gene IDs and HGNC symbols after enrichment analysis. The top 10 down-regulated and up-regulated genes are showed in Table 5 and Table 6. The list of DEGs were used for pathway analysis and active subnetwork analysis.

**Table 5. Top ten significant down-regulated genes in tumor samples of 55 LUAD patients**

| ensembl_gene_id | entrezgene | hgnc_symbol | logFC | adj.P.Val |
|---|---|---|---|---|
| ENSG00000182010 | 219790 | RTKN2 | -4.52455117194123 | 1.07397390772473e-42 |
| ENSG00000158764 | 142683 | ITLN2 | -7.4364942528429 | 3.19924465283634e-41 |
| ENSG00000102683 | 6445 | SGCG | -4.10485571819757 | 4.07515928515459e-41 |
| ENSG00000198873 | 2869 | GRK5 | -2.65790712992412 | 4.07515928515459e-41 |
| ENSG00000107742 | 9806 | SPOCK2 | -3.56967403596283 | 3.85300139768808e-40 |
| ENSG00000170323 | 2167 | FABP4 | -5.72790493543673 | 1.03033381509032e-39 |
| ENSG00000135063 | 9413 | FAM189A2 | -3.53046742312343 | 1.03117504787973e-39 |
| ENSG00000186994 | 256949 | KANK3 | -3.1101996380779 | 1.15468325581686e-39 |
| ENSG00000150625 | 2823 | GPM6A | -5.17438700689996 | 1.5648953870669e-39 |
| ENSG00000154721 | 58494 | JAM2 | -2.50261146610761 | 1.92231892168565e-39 |

**Table 6. Top ten significant up-regulated genes in tumor samples of 55 LUAD patients**

| ensembl_gene_id | entrezgene | hgnc_symbol | logFC | adj.P.Val |
|---|---|---|---|---|
| ENSG00000183010 | 5831 | PYCR1 | 3.5139225242735 | 3.06017765569688e-41 |
| ENSG00000059573 | 5832 | ALDH18A1 | 1.68852856318992 | 6.30895314373162e-38 |
| ENSG00000164466 | 94081 | SFXN1 | 1.5322079314688 | 5.01920971916517e-37 |
| ENSG00000135052 | 51280 | GOLM1 | 2.51608337184892 | 1.73125209540521e-35 |
| ENSG00000180198 | 1104 | RCC1 | 1.62119814668367 | 1.82637777402036e-34 |
| ENSG00000155660 | 9601 | PDIA4 | 1.6848754492746 | 2.37855372052335e-34 |
| ENSG00000096063 | 6732 | SRPK1 | 1.62823462104507 | 2.66740561460568e-34 |
| ENSG00000128050 | 10606 | PAICS | 1.65390171937903 | 4.22169230063646e-31 |
| ENSG00000111344 | 8437 | RASAL1 | 3.57173273242386 | 1.08251787193746e-30 |
| ENSG00000173457 | 26472 | PPP1R14B | 1.86684316566064 | 7.07845976872399e-30 |

## Active Subnetwork and Pathway Analysis

The output of Differentially Expression Analysis (DEA) containing differentially expressed genes with their Ensembl IDs and adjusted p-values (q-values) were used as input of *DEsubs* R package. The active subnetworks of differentially expressed genes in tumor samples of both 55 LUAD patients were determined by *DEsubs* package and results were represented as graphs at subnetwork and organism levels. *DEsubs* package identified 35 subnetworks including 192 genes, 14 of them including more than three genes, 8 of them including three genes and the others including two genes. In Figure 6, the top ten significant genes which play role in determined subnetworks are represented with their q-values. These genes are FABP4, WNT3A, EDNRB, TEK, AGER, EPAS1, ACADL, PDIA4, ANGPT4, KL. In this analysis, 35 subnetworks were determined and the first three subnetworks are presented in Figure 7, 8 and 9. When we look at the subnetworks' graphs, in subnetwork 1 (Figure 7), the prominent genes are WNT genes which are members of WNT pathway, a major evolutionary conserved signaling pathway playing role in cell differentiation, cell migration and organogenesis during development and highly related to lung cancer; in subnetwork 3 (Figure 9), the prominent gene is AKT3 which is one of the AKT family members which play role in tumorigenesis and are modulators of several tumors. The pathways of subnetwork genes are mostly cancer related pathways such as melanoma, glioma, colorectal cancer, chronic myeloid leukemia, basal cell carcinoma, apoptosis, erbb signaling, jak-stat signaling and map kinase signaling pathways (Figure 10).

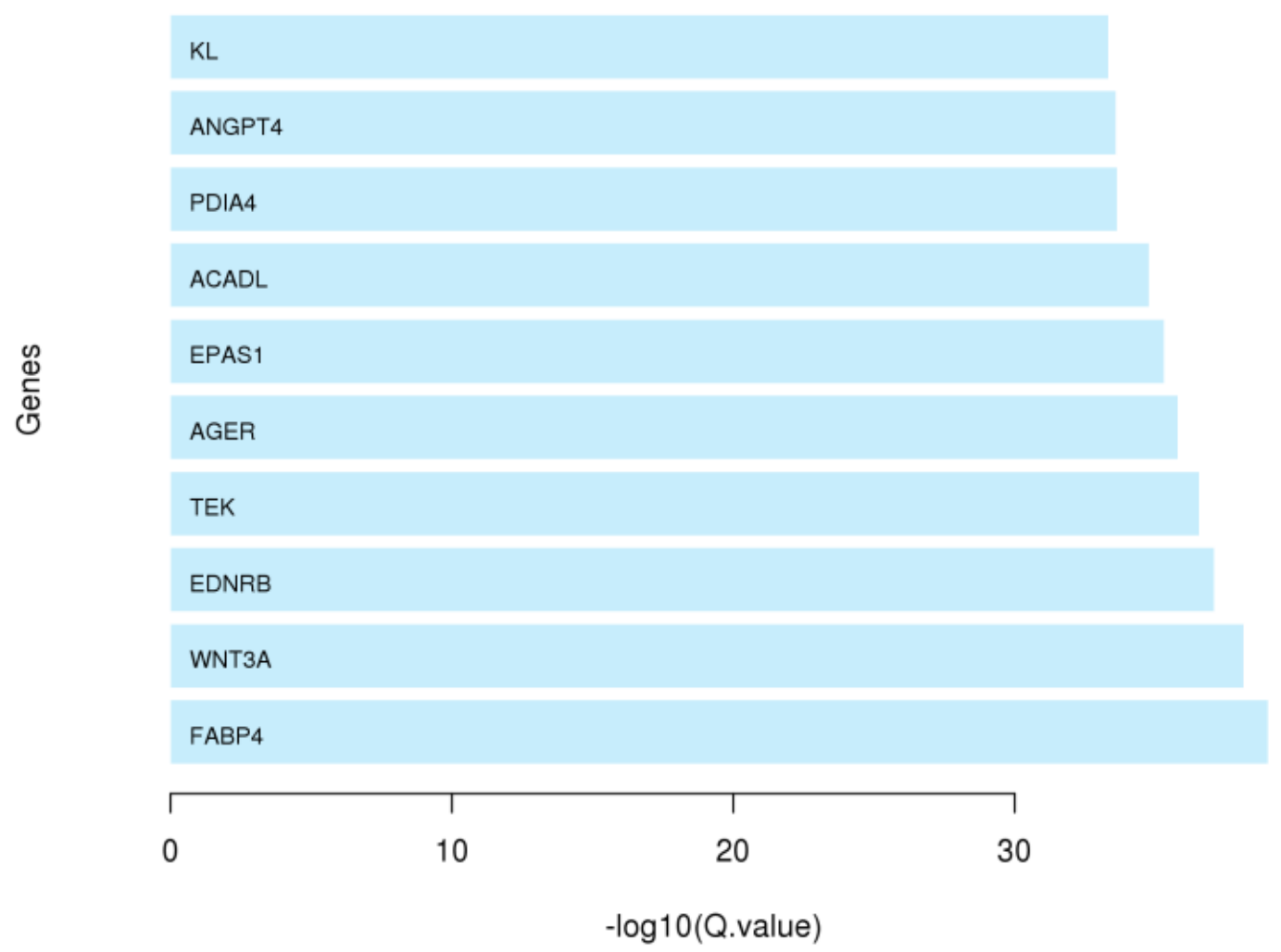

**Figure 6. Top 10 significant subnetwork genes in tumor samples of 55 LUAD patients**

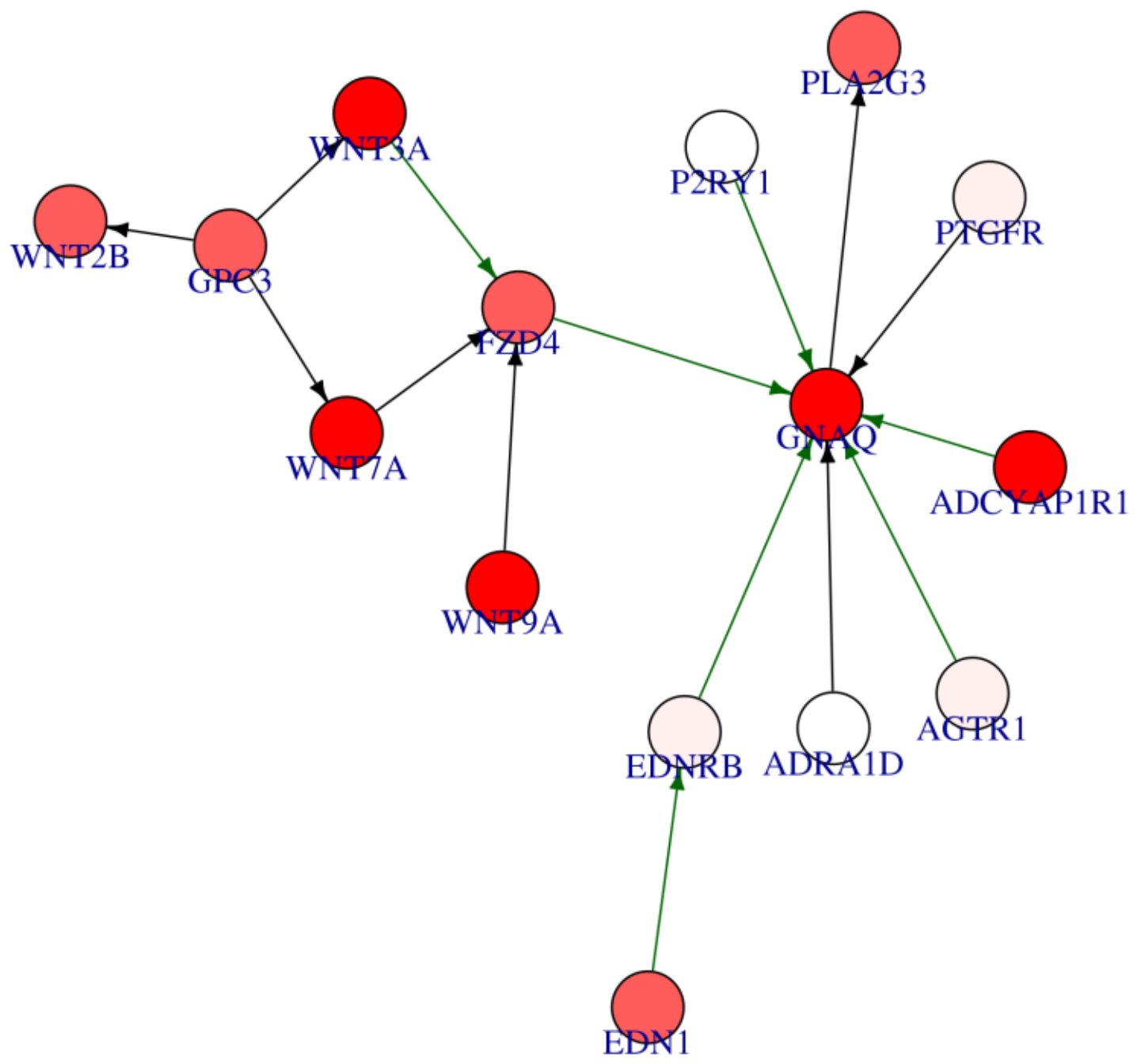

**Figure 7. Graph of subnetwork 1 from 55 LUAD patients**

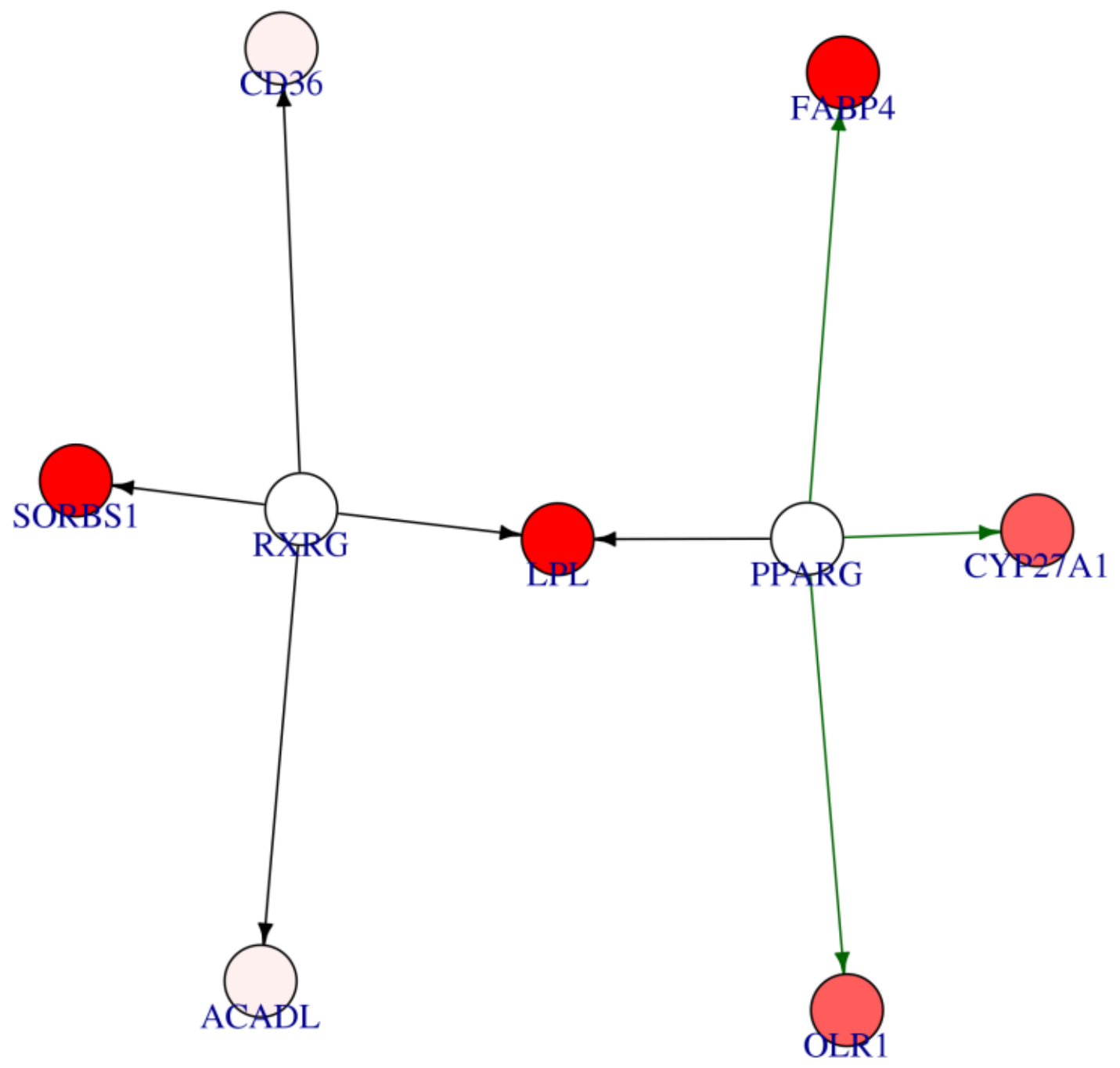

**Figure 8. Graph of subnetwork 2 from 55 LUAD patients**

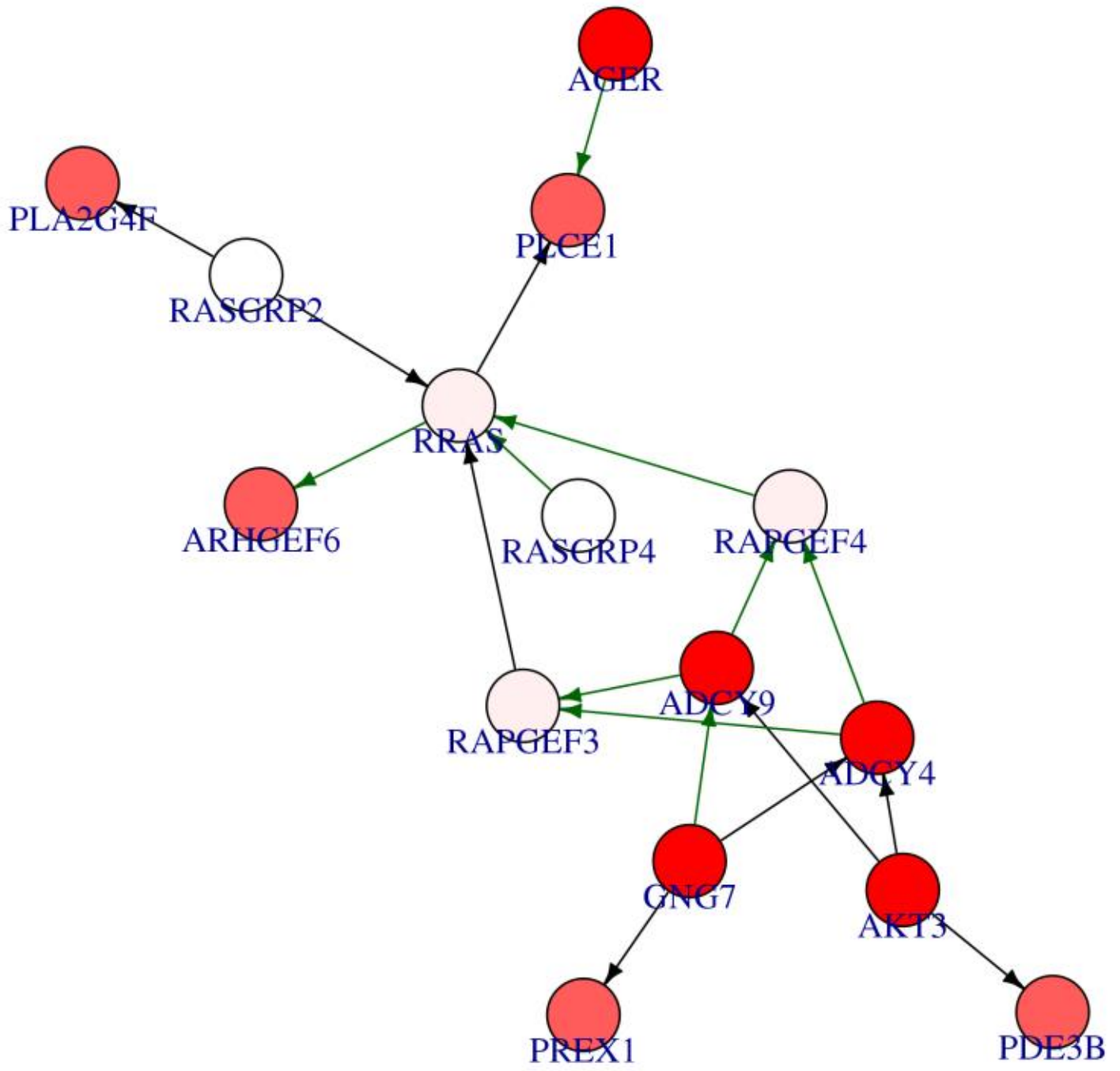

**Figure 9. Graph of subnetwork 3 from 55 LUAD patients**

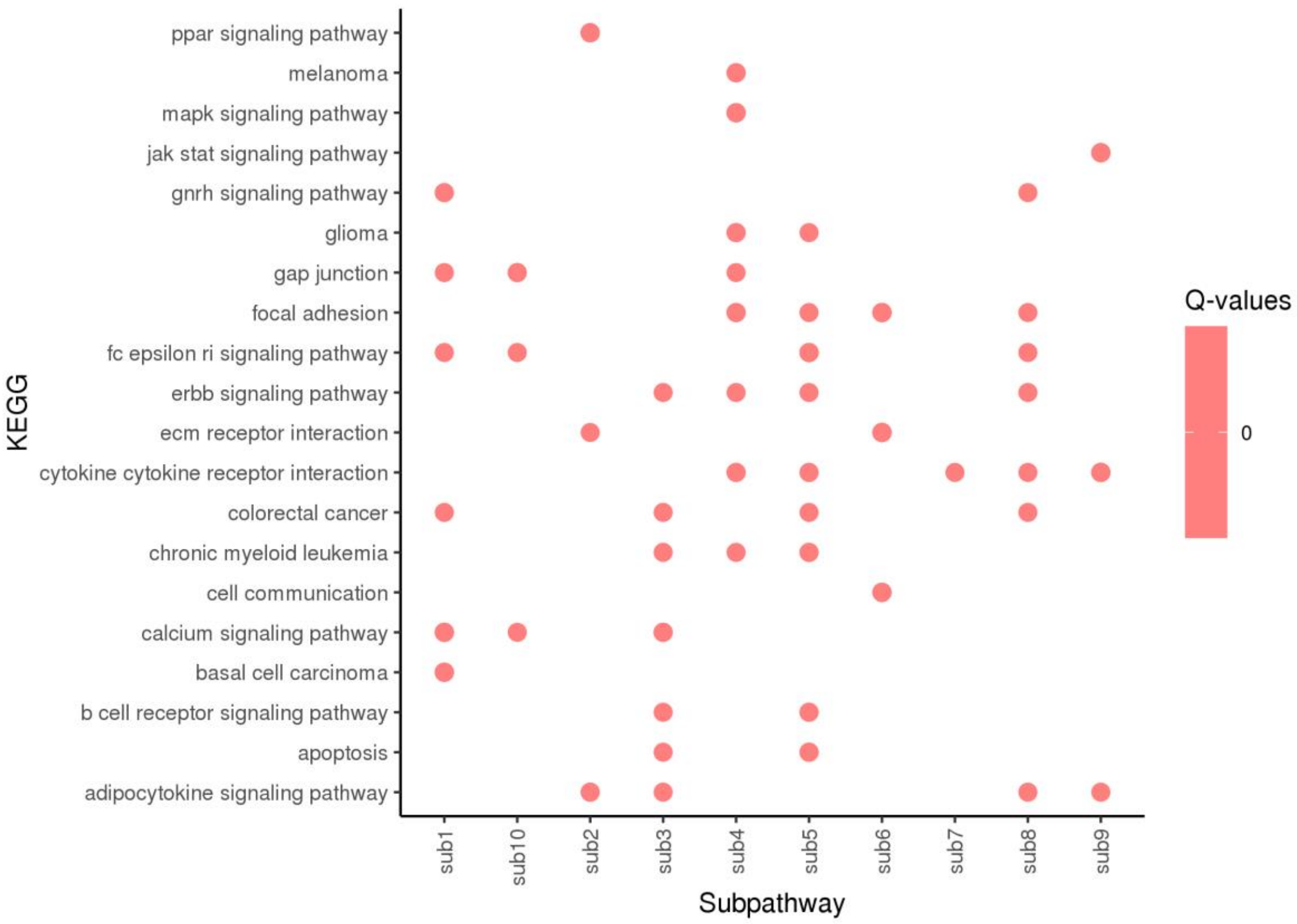

**Figure 10. KEGG pathways which includes active subnetwork genes of 55 patients with LUAD**

## Statistical Analysis

In order to identify a molecular prognosis risk model, the clinical data of all patients in TCGA LUAD project (Table 7) was downloaded by *TCGAbiolinks* R package and separated as training data of 55 LUAD patients who have paired samples for *RNAseq* data and used for gene signature construction; and test data of remaining 422 LUAD patients after removing patients who have missing values in clinical data. Different gene signatures were generated from the genes which have prognostic ability. The univariate cox regression analysis was performed for significant SNV genes, significant CNV genes, significant DEGs and active subnetwork DEGs in tumor samples of 55 patients with LUAD. There were 38 CNV genes, 463 DEGs and 37 subnetwork DEGs (DEsubs) with prognostic ability after univariate analysis and logRank test ($p < 0.05$). SNV genes did not have significant prognostic ability. Then different data categories (DEGs; DEsubs; CNVs; CNVs + DEGs, CNVs + DEsubs; CNVs + DEGs + SNVs; CNVs + DEsubs + SNVs) were generated by using significant prognostic genes. These data categories underwent the Cox proportional hazards regression with Lasso penalty and LOOCV. Gene models from different categories were generated by using *glmnet* R package which gives active genes with their coefficients. The genes in the models were DEPTOR, ZBTB16, BCHE, MGLL, MASP2, TNNI2, RAPGEF3, SGK2, MYO1A, CYP24A1, PODXL2, CCNA1 from DEGs category; THRA, RAPGEF3, LAMB2 from DEsubs category; SNX13, AC080080.1, RNMTL1P2, AC080080.2 from CNVs category; THRA, RAPGEF3, LAMB2 from CNVs + DEsubs. The genes in CNVs + DEGs and CNVs + DEGs + SNVs categories were same with the genes in DEGs category; the genes in CNVs + DEsubs + SNVs were in the CNVs + DEsubs category. Then, c-index analysis was performed to identify the survival predictive ability of the gene models identified from different categories in Figure 11. The higher c-index score was 0.858 from DEGs gene model. This gene model was chosen as best candidate prognosis gene signature for LUAD.

**Table 7. Summary of clinical features of 55 and 510 patients with LUAD**

| Category | Number | |
|---|---|---|
| | 55 patients | 510 patients |
| Age at diagnosis (median; range) | 66 (42-86) | 66 (33-88) |
| Gender | | |
| Female | 33 | 273 |
| Male | 22 | 237 |
| Tumor stage | | |
| I | 28 | 275 |
| II | 12 | 119 |
| III | 12 | 84 |
| IV | 2 | 25 |
| NA | 1 | 7 |
| Vital status | | |
| Alive | 31 | 326 |
| Dead | 24 | 184 |

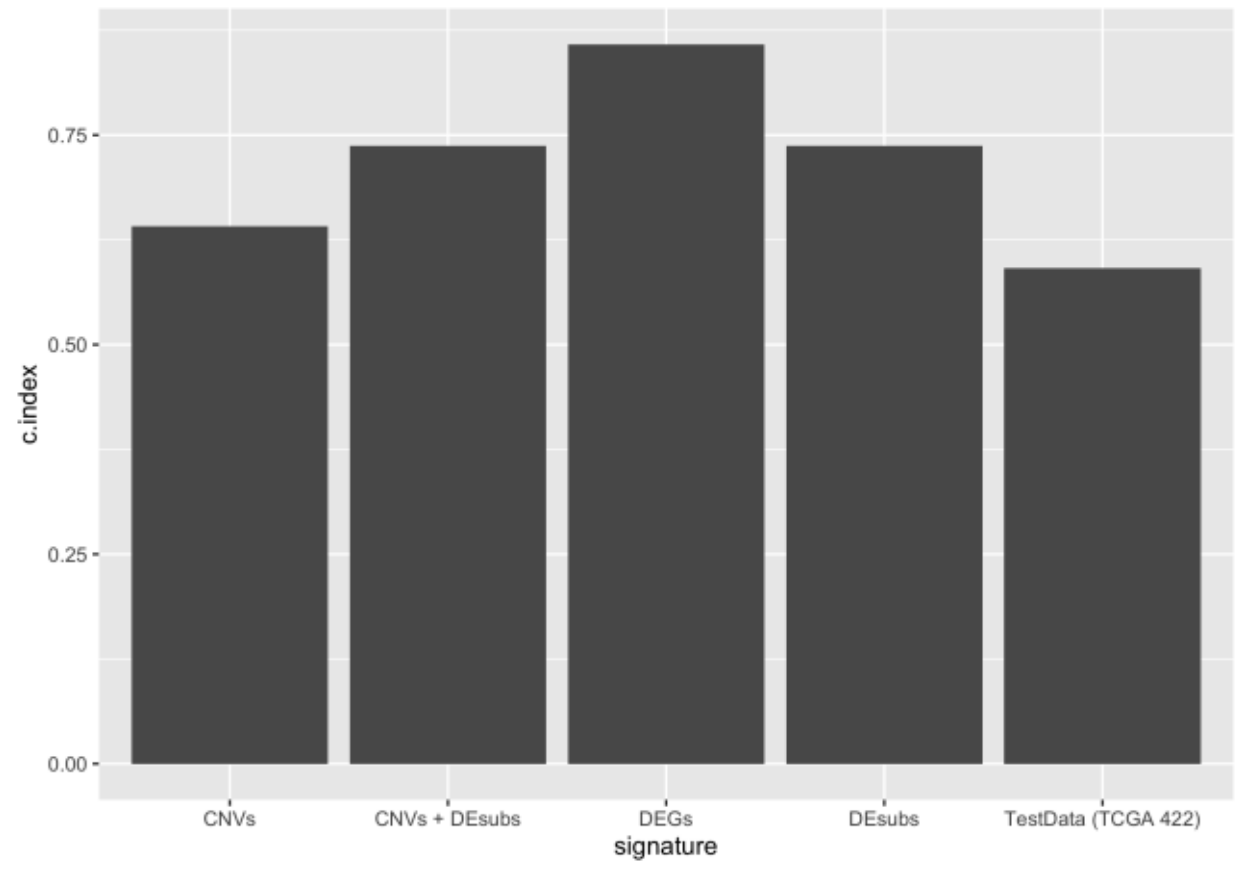

**Figure 11. The c-index of different gene categories in training data and selected signature in testing data.**

Multivariate Cox regression analysis was performed for the genes in the chosen gene signature and risk scores of each patient in training data (55 LUAD patients) were calculated by using coefficient values and normalized expression values ($log_2$+1) in tumor samples. Then the patients were clustered into high-risk and low-risk groups by using maxstat (maximally selected rank statistics) method based on optimal cutpoints for numerical variables by using *survminer* R package. When we performed Kaplan-Meier (KM) survival analysis to demonstrate the overall survival of risk groups stratified based on gene signature, patients with high-risk score demonstrated poor overall survival ($p < 0.0001$) than those with low-risk score in training dataset (Figure 12).

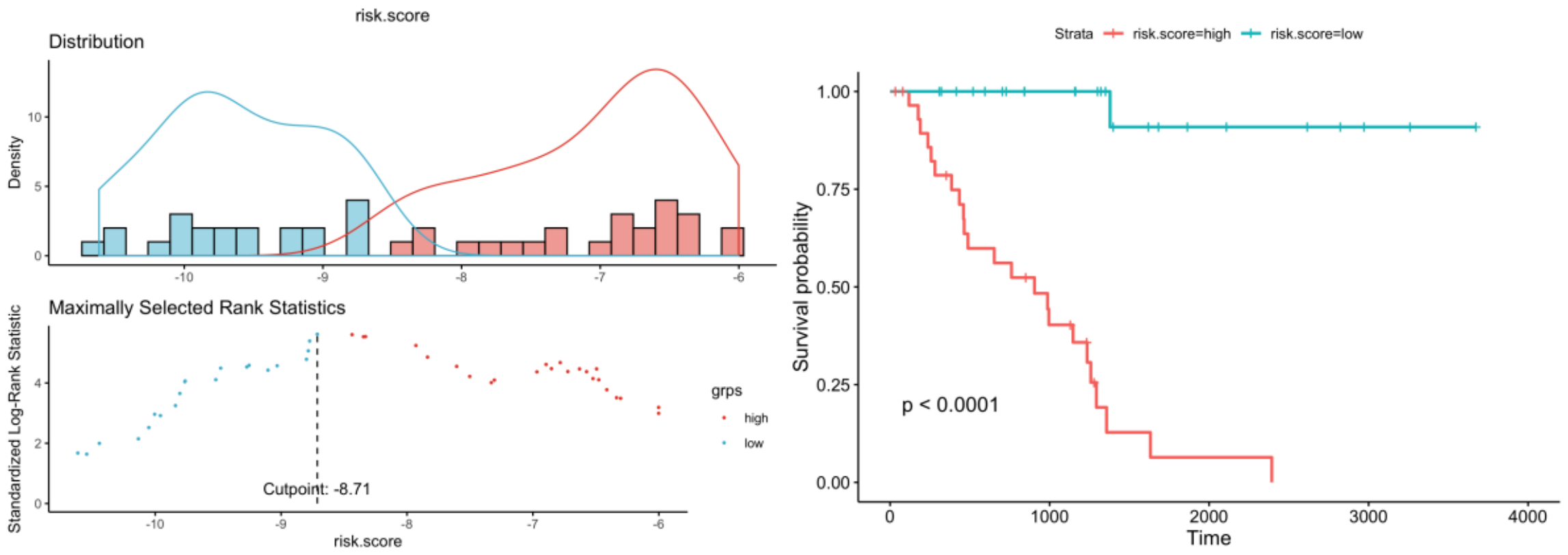

**Figure 12. Risk clustering and Kaplan-Meier survival analysis of the signature for training data**

The ROC curve analysis was performed to compare sensitivity and specificity of the predictive ability of risk score based on chosen gene signature. AUC values were 0.883 for 1-year, 0.813 for 2-year, 0.943 for 5-year and 0.976 for 10-year survival prediction (Figure 13a). These high AUC values showed that the risk scores calculated based on chosen gene signature can highly predict the overall survival.

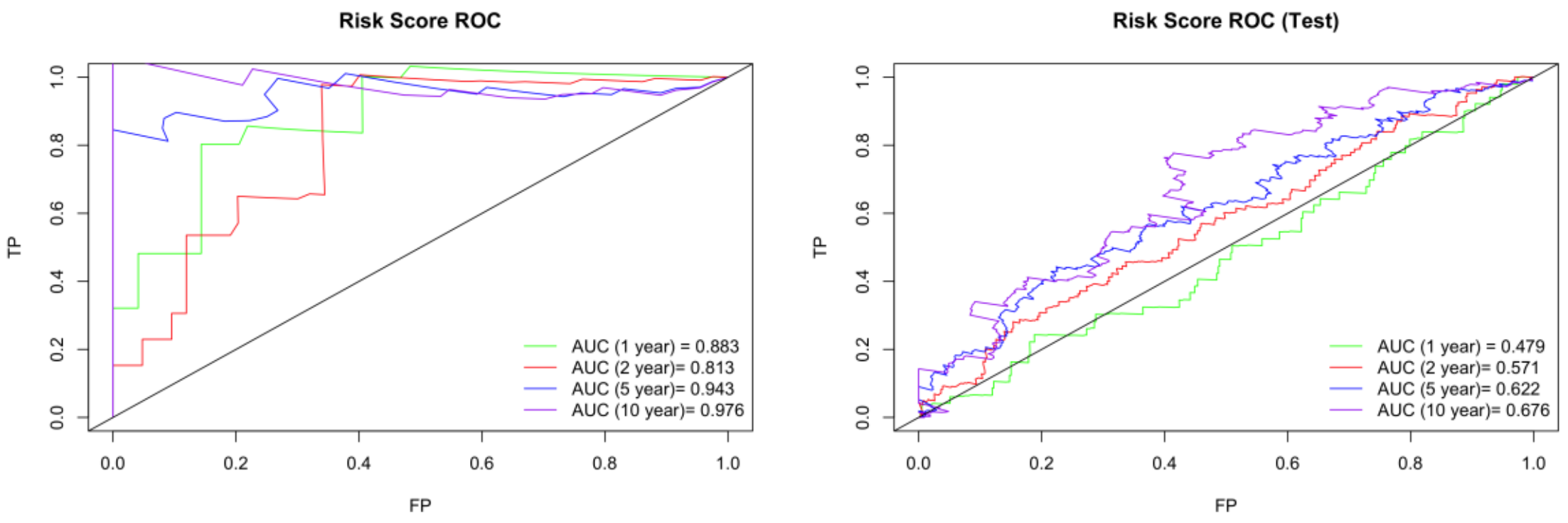

**Figure 13. ROC curve analysis for 1, 2, 5 and 10-year survival prediction by the signature in both training and test data**

When we performed the correlation analysis between tumor stages, mutation counts and gene expressions of signature genes, there was a significant difference of tumor stages between risk groups although there was no difference of total SNV mutation count between groups (Figure 14). However, as expected gene expression levels were significantly different between high-risk and low-risk groups in training data (55 LUAD patients) (Figure 15). The expression levels of the BCHE, DEPTOR, MASP2, MGLL, MYO1A, PODXL2, RAPGEF3, SGK2, TNNI2, and ZBTB16, genes were lower in high-risk group while the expression levels of the CCNA1 and CYP24A1 genes were higher in high-risk group (Figure 15).

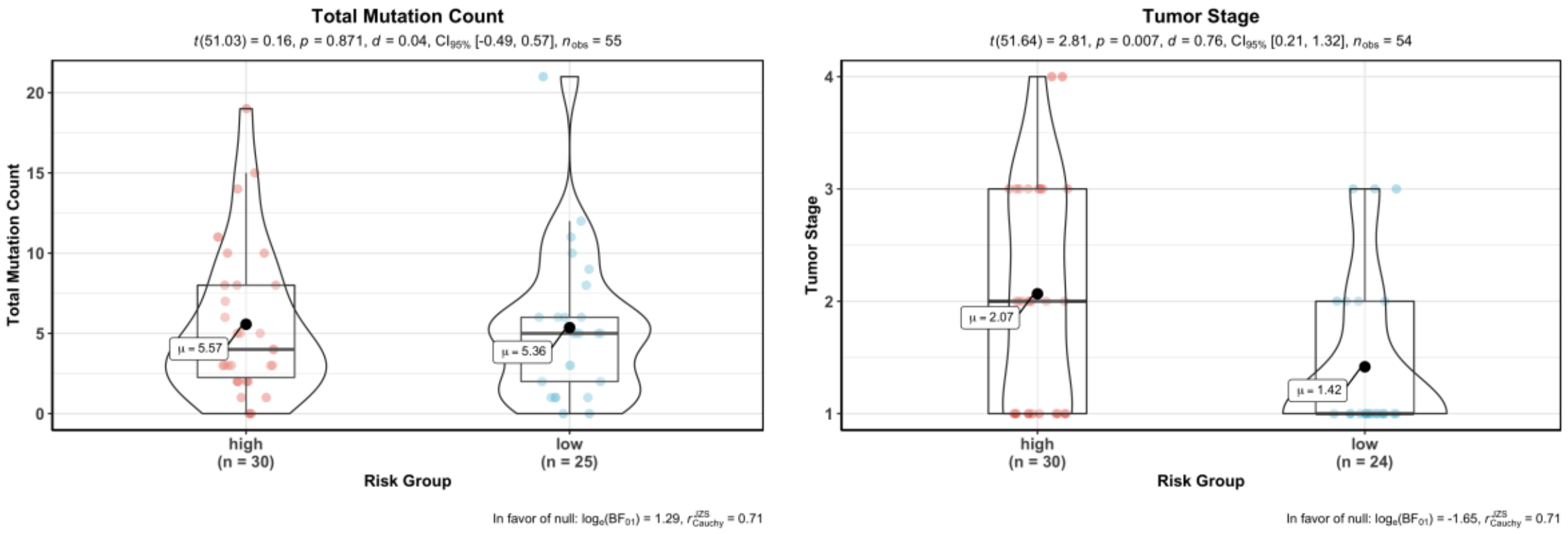

**Figure 14. Correlation analysis between risk groups and total mutation count and tumor stage in tumor samples of training data**

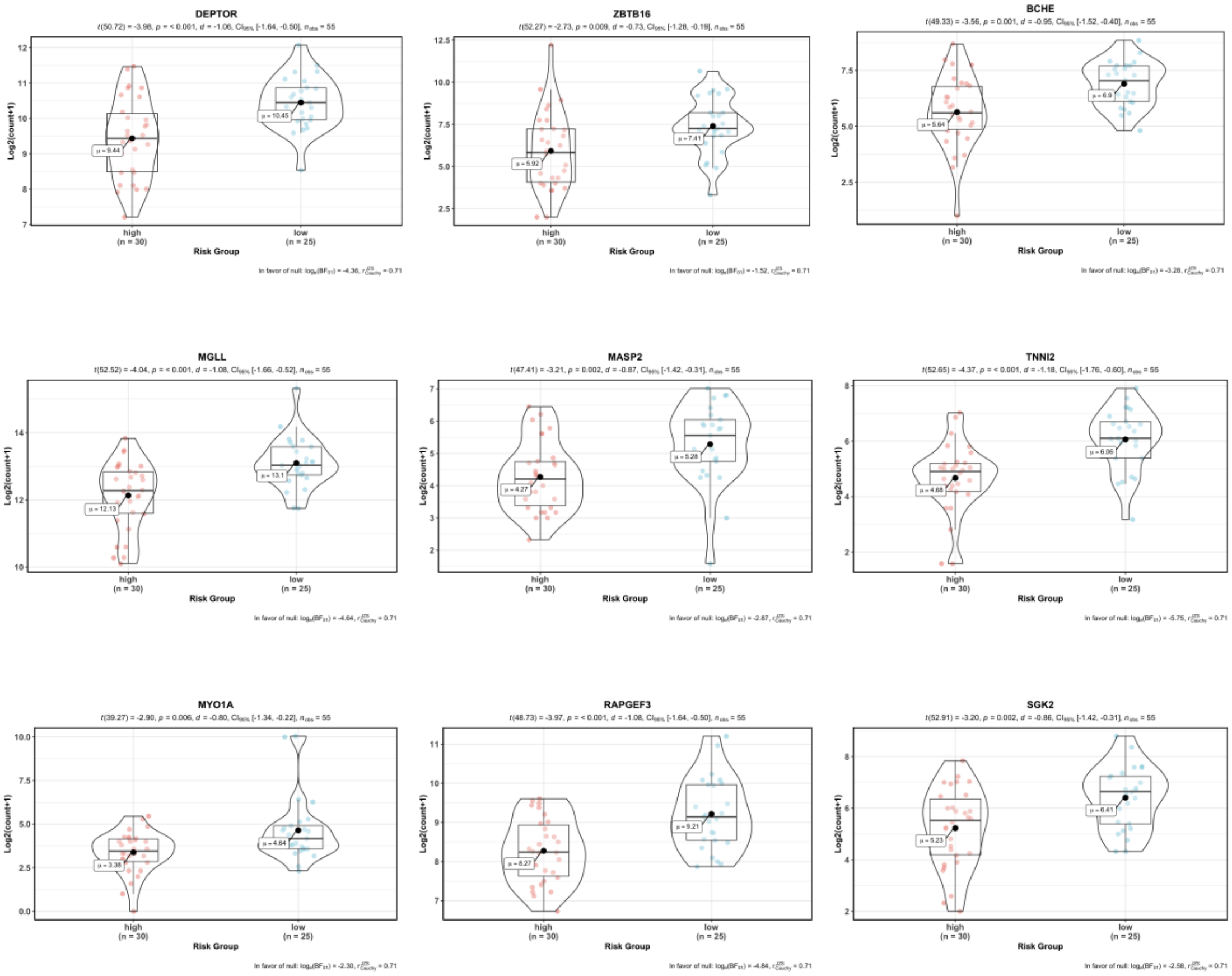

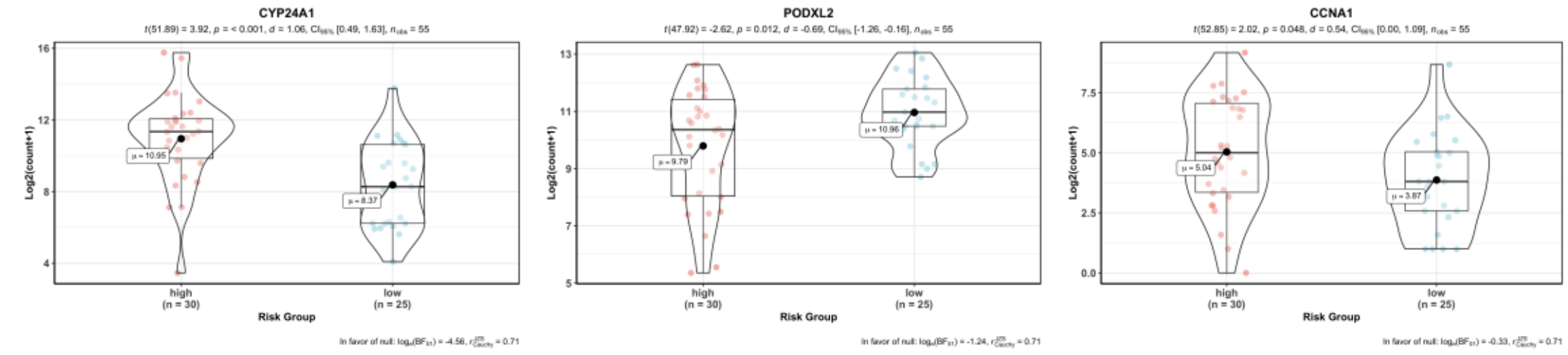

**Figure 15. Violin plot showing the expression levels of the signature genes between low-risk and high-risk groups in tumor samples of training data**

In order to validate our signature, we calculated c-index for the prediction of overall survival of the 442 TCGA patients with LUAD (test data) and the c-index was 0.591 which is lower than the c-index of training data (0.858). Then, multivariate cox regression analysis was performed for the signature genes in test data. The risk score for each patient was calculated by adding the multiplication of normalized gene expression level in tumor samples and multivariate cox regression coefficient value of each gene in signature. Patients in test dataset were divided into high-risk and low-risk groups by using maxstat (maximally selected rank statistics) method from using *survminer* R package (Figure 16a). Patients in high-risk group had poor overall survival significantly ($p < 0.00055$). The ROC curve analysis was performed to compare sensitivity and specificity of the predictive ability of risk score in the test dataset. AUC values were 0.479 for 1-year, 0.571 for 2-year, 0.622 for 5-year and 0.676 for 10-year survival prediction (Figure 13b). The AUC values of risk scores calculated based on chosen gene signature were very low according to the AUC values of training data. Although the survival predictive ability (c-index) of our gene signature and AUC values of risk score in test data was low, our 12-gene signature could separate patients into two groups which have significant overall survival difference (Figure 16b).

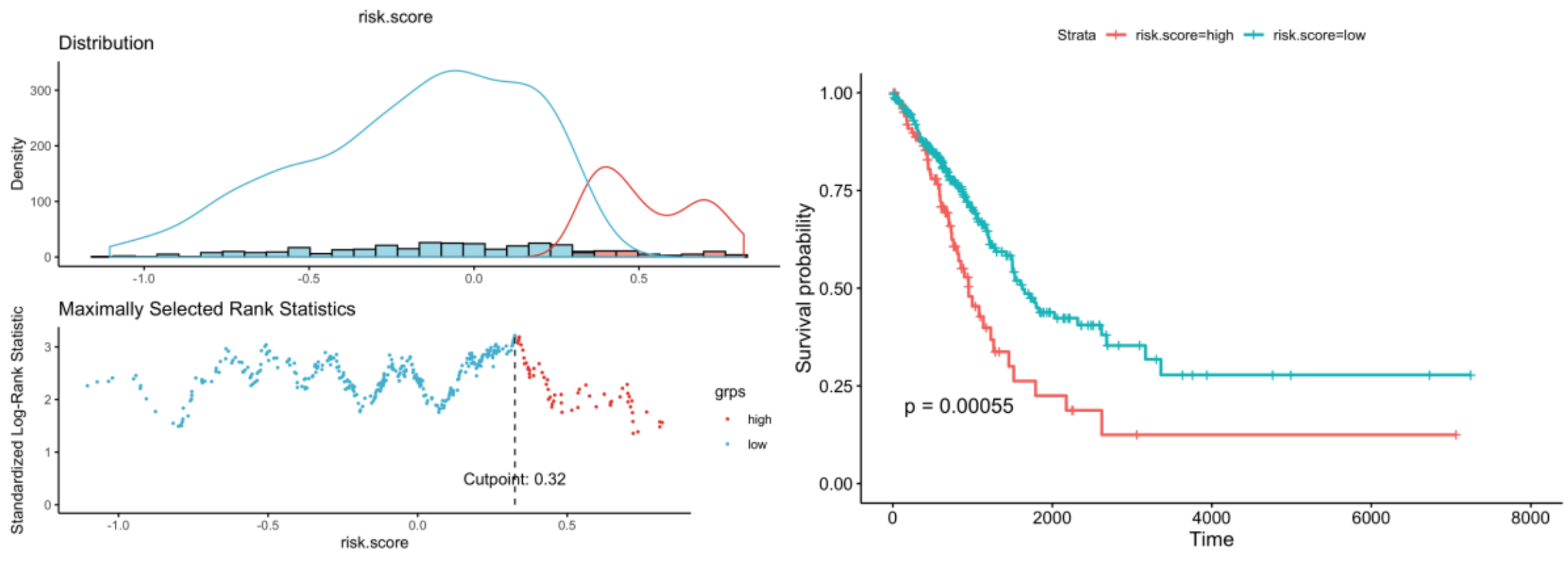

**Figure 16. Risk clustering and Kaplan-Meier survival analysis of the signature for test data**

We performed the correlation analysis between tumor stages, mutation counts and gene expressions of signature genes for test data, there was a slight significant difference of tumor stages between risk groups although there was no difference of total SNV mutation count between groups (Figure 17). The gene expression levels of 6 signature genes (BCHE, CCNA1, DEPTOR, MASP2, MGLL, TNNI2) were significantly different between high-risk and low-risk groups however, the gene expression levels of other 6 signature genes (CYP24A1, MYO1A, PODXL2, RAPGEF3, SGK2, ZBTB16) do not have significant difference in test data. The expression levels of the CCNA1 and TNNI2 genes were lower in high-risk group while the expression levels of the BCHE, DEPTOR, MASP2 and MGLL genes were higher in high-risk group (Figure 18).

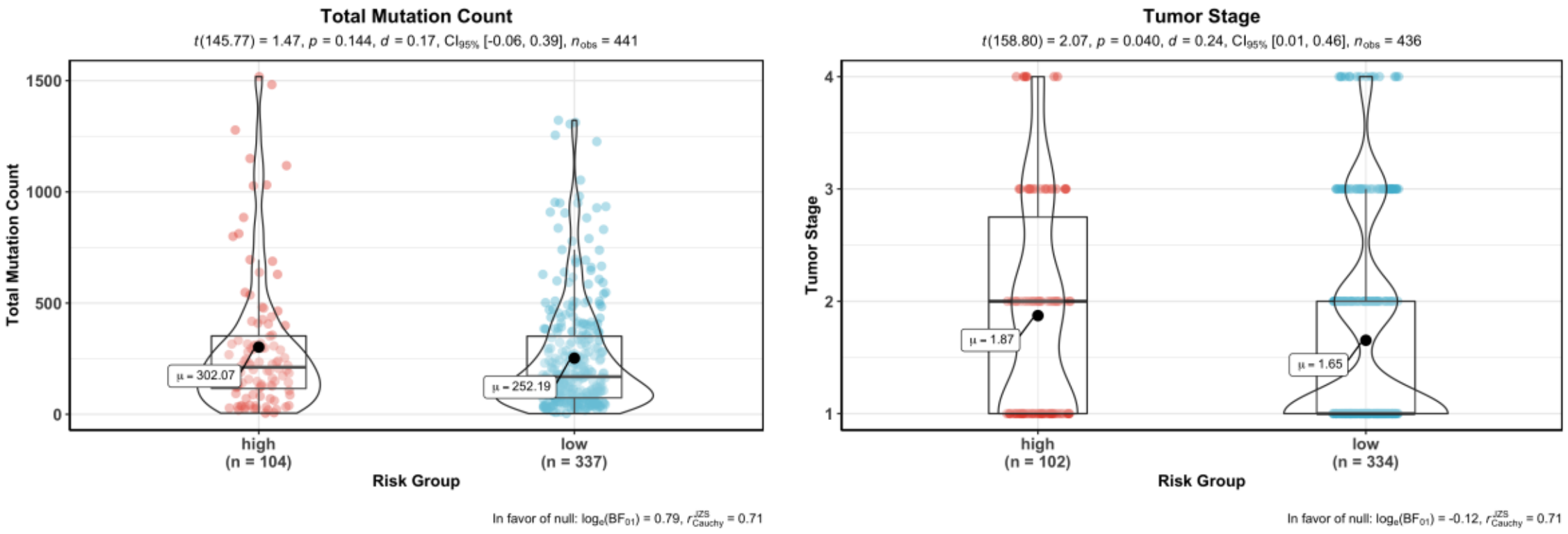

**Figure 17. Correlation analysis between risk groups and total mutation count and tumor stage in tumor samples of test data**

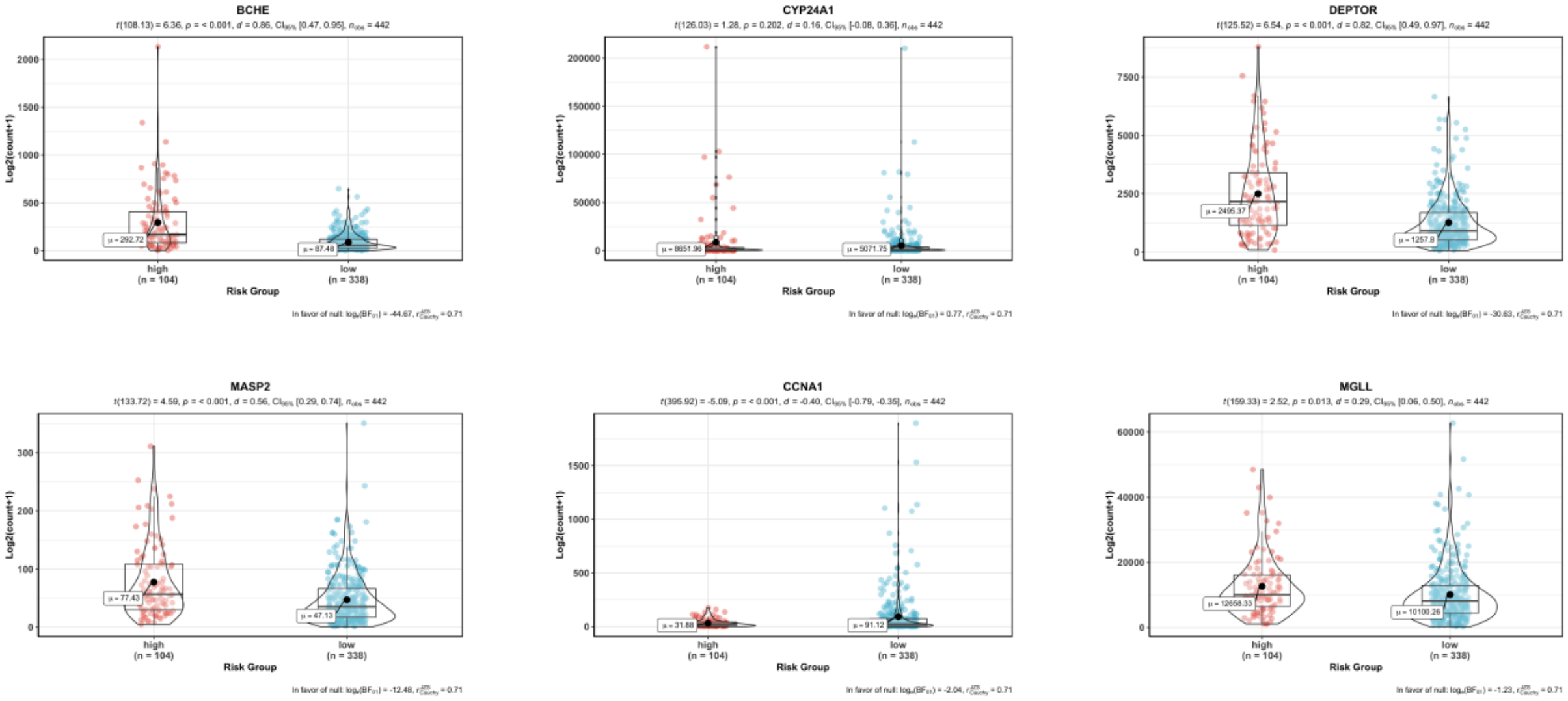

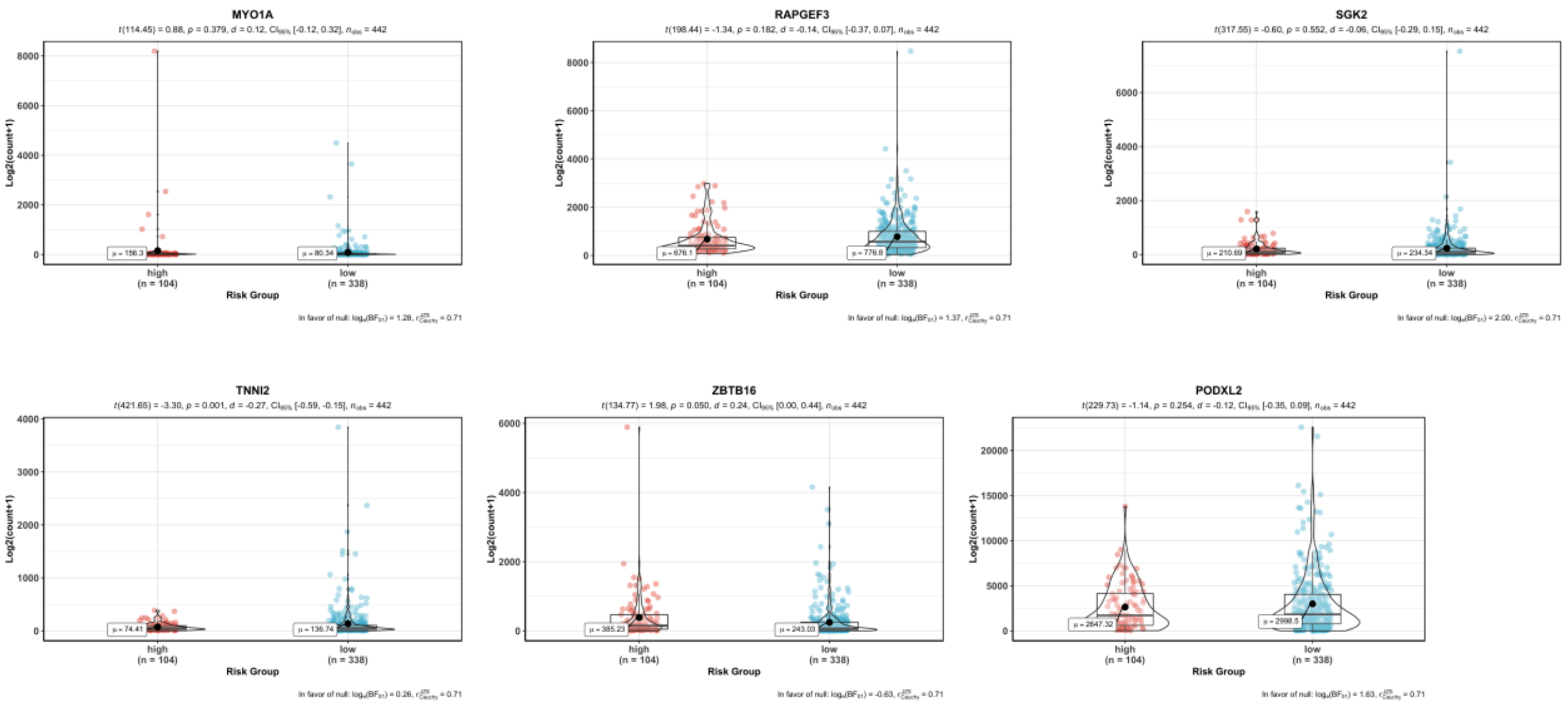

**Figure 18. Violin plot showing the expression levels of the signature genes between low-risk and high-risk groups in tumor samples of test data**

## Discussion

Lung adenocarcinoma (LUAD) is the most common form of lung cancer which is most common cancer and responsible for largest number of deaths worldwide. In order to characterize genomic and transcriptomic abnormalities of lung cancer and to determine clinical status of patients, integrative analyses have been performed by using different types of molecular data. Recently, prognosis risk signatures have been generated to cluster patients with lung adenocarcinoma. However, mostly gene expression data has been used for this purpose.

In this study, we performed an integrative analysis by using level-3 data of SNVs, CNVs and RNAseq data of patients with lung adenocarcinoma in TCGA project. We aimed to identify genomic and transcriptomic abnormalities that might be used to generate a molecular signature. We determined significant mutated genes; amplified and deleted genes; and differentially expressed genes (DEGs) significantly and their active subnetworks by using R packages. Then we performed univariate and multivariate Cox Proportional Hazards Regression (CPHR) analysis with LOOCV and Lasso penalty to identify predictor genes on survival time of patients with lung adenocarcinoma.

We identified 6 and 82 mutated genes which are candidate driver genes in tumor samples of 55 LUAD patients and those of 510 LUAD patients, respectively. KRAS and EGFR oncogenes with TP53, STK11, RB1 and MGA tumor suppressors were mutated significantly in small cohort of patients. The mutated 82 genes of big cohort of patients include the 6 genes above and also previously identified lung adenocarcinoma related genes such as KRAS, TP53, STK11, RB1, NF1, RMB10, BRAF, KEAP1, CDKN2A, SETD2, ARID1A, SMARCA4 and MGA (The Cancer Genome Atlas Research Network, 2014); EGFR and ERBB2 (The Cancer Genome Atlas Research Network, 2014; Berger, 2016); and PIK3CA (The Cancer Genome Atlas Research Network, 2014; Deng, 2017). Besides, MAP2K1 and MAP2K4 mutations can be related with MAPK pathway activity as identified in TCGA lung adenocarcinoma original article (The Cancer Genome Atlas Research Network, 2014). Loss-of-function MGA mutations with MYC amplification in lung adenocarcinoma have been newly described (The Cancer Genome Atlas Research Network, 2014) and MGA gene was identified by *SomInaClust* analysis in our study. MGA, encodes MAX gene-associated protein which is a MYC-interacting transcription factor and antagonizes the transcriptional regulation of MYC involved in cancer processes (Romero, 2014).

We identified amplified and deleted genes which have copy number variations in tumor samples of patients with lung adenocarcinoma. We identified significant copy number altered genes which play role in immune system pathways, metabolism pathways with small cell lung cancer pathway and molecular mechanism of cancer pathway. We analyzed differentially gene expression in tumor samples compared to paired normal samples of 55 patients with lung adenocarcinoma and 3575 genes were dysregulated more than two-fold, significantly (q-value < 0.01). The upregulated genes mostly play role in cell cycle and proliferation pathways such as G2/M damage checkpoint regulation, cell cycle control of chromosomal replication, ATM signaling, hereditary breast cancer signaling, bladder cancer signaling and HIF1 signaling pathways. The downregulated genes play role in cAMP-mediated signaling, g-protein coupled receptor signaling, Gαi signaling and other immune system pathways such as complement system, granulocyte/agranulocyte adhesion and diapedesis, dendritic cell maturation and T helper cell differentiation. Then we determined differentially expressed genes (DEGs) in active subnetworks of PPI network in tumor samples and we identified 192 DEGs in 35 subnetworks. These genes play role in mostly cancer related pathways such as melanoma, glioma, colorectal cancer, chronic myeloid leukemia, basal cell carcinoma, apoptosis, erbb signaling, jak-stat signaling and map kinase signaling pathways (Figure 10).

We integrated significant SNVs, CNVs, DEGs and DEGs in active subnetworks by performing multivariate Cox Proportional Hazards Regression (CPHR) analysis with LOOCV and Lasso penalty after univariate CPHR, we determined 12-gene expression signature (BCHE, CCNA1, CYP24A1, DEPTOR, MASP2, MGLL, MYO1A, PODXL2, RAPGEF3, SGK2, TNNI2, ZBTB16) which has 0.858 and 0.591 c-index score for training and test data, respectively. Moreover, this 12-gene expression signature had 0.883, 0.813, 0.943 and 0.976 AUC values for 1, 2, 5 and 10-year survival prediction, respectively, for training data. Same 12-gene expression signature had 0.479, 0.571, 0.622 and 0.676 AUC values for 1, 2, 5 and 10-year survival prediction, respectively, for test data. We clustered the patients for both training and test analysis, into high-risk and low-risk group based on risk scores calculated by using expression levels and multivariate CPHR coefficients of 12 genes in signature. Kaplan-Meier survival analysis showed highly significant overall survival difference between high-risk and low-risk groups for both training data ($p < 0.0001$) and test data ($p = 0.00055$).

All genes in 12-gene signature are cancer-related and play role in lung cancer pathways which are candidates of molecular targeting. BCHE (Butyrylcholinesterase) activity in lung adenocarcinoma is less than in adjacent non-cancerous tissue (Martinez-Moreno P. 2006); and BCHE is one of two potential diagnostic markers in plasma/serum for non-small cell lung cancer (Shin J. 2017). CCNA1 (Cyclin A1) is a cell cycle regulator protein and was down-regulated in non-small cell lung cancer and CCNA1 promoter was hypermethylated in 70% of lung tumors which has wild-type p53, but was not methylated in cells with mutant p53 (David S. Shames, 2006). CCNA1 plays a role in p53-mediated G2 cell cycle arrest and apoptosis in non-small cell lung cancer cells and upregulation of cyclin A1 resulted in apoptosis (Rivera A. 2006). However, Cho et al. determined that knock-down of CCNA1 using siRNA, induced apoptosis in non-small cell lung cancer cells (Cho N. H. 2006). CYP24A1 expression level was highly increased in lung adenocarcinoma compared to normal lung tissue samples and CYP24A1 overexpression was associated with poorer survival, increased cell growth and invasion, and increased RAS protein expression in lung adenocarcinoma (Guoan Chen, 2010; Hiroe Shiratsuchi, 2016; NAN GE, 2017; Meng Li, 2019). Knockdown of CYP24A1 significantly decreased cell proliferation resulted in tumor growth delay and smaller tumor size with decreased RAS protein level, thus reducing phosphorylated AKT (Hiroe Shiratsuchi, 2016). DEPTOR (DEP domain-containing mTOR-interacting protein), a natural mTOR inhibitor, was downregulated by activation of EGFR signaling. EGFR inhibition by Gefitinib resulted DEPTOR accumulation. DEPTOR inhibited proliferation, migration, invasion and the tumor growth of lung adenocarcinoma. DEPTOR induction inhibited EGFR mediated tumor progression (Xuefeng Zhou, 2016). DEPTOR depletion can induce EMT in cancer cells and DEPTOR plays a critical role in EMT regulation by BMK1 (Runqiang Chen, 2012). DEPTOR was also identified as one of the 77 clinically relevant predictive biomarker at TGFβ-EMT signature generated by microarray analysis of TGFβ-1 treated non-small cell lung cancer cells. TGFβ-EMT gene signature could predicted overall survival and metastasis-free survival in lung adenocarcinoma (Edna Gordian, 2019). MASP-2 (Mannan-binding lectin-associated serine protease 2) is a plasma protein involved in lectin pathway of complement system which promotes cell dedifferentiation, proliferation, migration and reduced apoptosis. Complement activation in the tumor microenvironment enhances tumor growth and increases metastasis (Vahid Afshar-Kharghan, 2017). High MASP-2 levels concentration in serum significantly correlated with recurrent cancer disease and with poor survival, thus the MASP-2 level had an independent prognostic value in the patients (Henriette Ytting, 2005). MBL/MASP complex activity was significantly increased in patients with colorectal cancer, too (H. Ytting, 2004). MGLL (Monoglyceride lipase) is highly expressed in aggressive human cancer cells and primary tumors, where it regulates a fatty acid network enriched in oncogenic signaling lipids that promotes migration, invasion, survival, and in vivo tumor growth (Daniel K. Nomura, 2010). MGLL expression was significantly reduced in the majority of primary human lung cancers and primary colorectal cancers compared to normal tissues (Renyan Liu, 2018; H Sun, 2013). MGL suppressed colony formation in tumor cell lines and knockdown of MGL resulted in increased Akt phosphorylation. MGL plays a negative regulatory role in phosphatidylinositol-3 kinase/Akt signaling and tumor cell growth (H Sun, 2013). MGLL knock-out mice exhibited a higher incidence of neoplasia in lung (Renyan Liu, 2018). MYO1A (Myosin I a) expression was higher in ever smokers than in never smokers (Giulia Pintarelli, 2019). MYO1A had mutations and promoter hypermethylation in patients with colorectal cancer and gastric tumors; therefore, lower levels of MYO1A expression was associated with faster tumor progress and poor prognosis (Mazzolini 2012; Mazzolini 2013). Podocalyxin is an anti-adhesive transmembrane protein played role in the development of more aggressive breast and prostate cancer (9, 12 in Steven Sizemore, 2007). Podocalyxin (including PODXL1, PODXL2 and PODXL3) induction resulted in altered migration and invasion, increased MMP expression with increased MAPK and PI3K activity through forming a complex with Ezrin protein, in breast and prostate cancer (Steven Sizemore, 2007). Mammalian exchange protein directly activated by cAMP isoform 1 (EPAC1), encoded by RAPGEF3 gene, acts as guanine exchange factor for Ras-like Rap small GTPases (Banerjee, 2015). EPAC1 expression was lower in lung cancer tissue compared to expression in normal specimens and associated with the degree malignancy and lymph-node metastasis (Qian Sun, 2018). SGK is one of three isoforms of the serum glucocorticoid regulated kinase family of serine/threonine kinases. SGK2 expression was upregulated in hepatocellular carcinoma and its downregulation inhibits cell migration and invasion (Junying Liu, 2017). Expression level of SGK1 was higher in squamous cell lung cancer and correlated with high grade tumors, tumors size and clinical stage (Claudia

Abbruzzese, 2012). Protein and mRNA expression of cardiac troponin I (TNNI3) were abnormally detected in non-small cell lung cancer tissues, lung adenocarcinoma cell line and lung squamous cell carcinoma cases while there was negative staining for TNNI3 in non-cancer lung tissues (Chao Chen, 2014). ZBTB16 (zinc finger and BTB domain containing 16), also known as the promyelocytic leukemia zinc finger protein (PLZF), was down-regulated in lymph node adenocarcinoma metastases and NSCLC samples by hypermethylation in the promoter region (Xiaotian Wang, 2013; Guang-Qian Xiao, 2015). Overexpression of ZBTB16 in lung cancer cell lines inhibited proliferation and increased apoptosis while the depletion of cytoplasmic PLZF was correlated with high tumor grade, lymph node metastasis, higher tumor stage and shorter overall survival (Xiaotian Wang, 2013; Guang-Qian Xiao, 2015). ZBTB16 was also down-regulated in never smoker patients with lung adenocarcinoma (YUNQIAN HU, 2015) and non-small cell lung cancer high-metastatic cell line compared with the low-metastatic cell line (RUIYING SUN, 2019).

## Conclusions

In this study we analyzed significant SNVs, CNVs and DEGs in active subnetworks, which have impact on overall survival of patients with lung adenocarcinoma in TCGA project. We determined 12-genes of which are strong candidates to be used as molecular signature for prediction of overall survival-based risk group of patients with lung adenocarcinoma. These genes can be used to cluster patients and determine the best candidates of drugs for the patient clusters which have different molecular nature.

These genes also have high potential for targeted cancer therapy of patients with lung adenocarcinoma.

## Methods

### Data

Simple Nucleotide Variation, Transcriptome Profiling, Copy Number Variation and Clinical data of both 55 LUAD patients who have paired (both normal and tumor samples) RNAseq data and of 510 patients who have all four types of data was downloaded from TCGA harmonized database by using R/Bioconductor TCGAbiolinks package (Colaprico, 2016). We analyzed the genomic alteration data including Simple Nucleotide Variations, Copy Number Variations; and transcriptomic variations from RNAseq data, processed using the reference of hg38; and clinical data of LUAD patients (Table 7).

### Identification of Significant Simple Nucleotide Variations

The Mutation Annotation Format (maf) file contained somatic mutations of all patients in TCGA LUAD project, was downloaded using TCGAbiolinks package. The other R/Bioconductor package, maftools (Mayakonda, 2018), were used to subset original maf file by tumor sample barcodes of patients of interest. Maftools package also summarizes the mutations and represents as summary plot and oncoplot. Significant mutated genes divided into two groups, oncogene (OG) or tumor suppressor gene (TSG), among tumor samples of 55 and 510 patients were identified seperately by using SomInaClust R package (Van den Eynden, 2015). SomInaClust works on the basic assumption that important genes in tumor samples have clustered on sequence and high number of inactivating mutations because of the selective pressure during tumorigenesis. Based on this assumption, oncogenes have clustered mutations, while tumor suppressors have inactivating (protein truncating) mutations. SomInaClust uses a reference step in which background mutation rate and hot spots are determined for genes existing in reference mutation database such as COSMIC database (v88) (Forbes, 2017).

### Identification of Significant Copy Number Variations

The TCGA LUAD CNV dataset for primary solid tumor samples, generated by Affymetrix Genome-Wide Human SNP Array 6.0 platform, was downloaded using TCGAbiolinks package. The significant aberrant genomic regions in tumor samples of 55 and 510 patients were identified separately by R/Bioconductor GAIA package (Morganella, 2011). NCBI IDs and Hugo Symbols of the genes with differential copy number were determined using biomaRt package (Durinck, 2009).

### Differential Expression Analysis (DEA)

The Transcriptome Profiling data in mRNA expression level (as unnormalized HTSeq raw counts) of 55 LUAD patients who have paired samples was downloaded by TCGABiolinks package. Differentially expressed genes were determined with FDR adjusted p-values (q-values) in tumor samples (TP) according to normal samples (NT) of 55 LUAD patients by limma-voom method using limma (Ritchie, 2015) and edgeR (McCarthy, 2012) R/Bioconductor packages. NCBI IDs and Hugo Symbols of the differentially expressed genes determined by the biomaRt R package.

### Active Subnetwork and Pathway Analysis

We identified the active subnetworks of differentially expressed genes in tumor samples of 55 LUAD patients using R/Bioconductor DEsubs package (Vrahatis, 2016). The output of limma package containing differentially expressed genes with their Ensembl IDs and FDR adjusted p-values (q-values) were used as input of DEsubs package. DEsubs package determines and represents the active subnetworks with their graphs both at subnetwork and pathway levels.

### Statistical Analysis

Clinical data of 55 and 510 patients was downloaded from TCGA database using the *TCGAbiolinks* package. Univariate Cox Proportional Hazards Regression analysis (Cox, 1972) and logRank test (Mantel, 1966) were performed using *survival* R package for significant SNV containing genes, significant CNV containing genes, DEGs and active subnetwork genes to identify genes with prognostic ability. For the genes with prognostic ability (p value < 0.05), Cox proportional hazards model (LOOCV) with Lasso penalty was used to identify best gene signature among different combinations of molecular levels (SNV genes, CNV genes, DEGs and active subnetwork genes) by using glmnet R package. Concordance index (c-index) was performed using *pec* R package to validate the predictive ability of different gene signatures. The larger c-index is used to determine the gene signature which has more accurate predictive ability. Multivariate cox proportional regression analysis was performed using *survival* R package for genes of selected signature and risk score of each patient was calculated using coefficient and expression values of the genes. Then, patients were clustered into high-risk group and low-risk group and Kaplan-Meier (KM) survival curves (Kaplan &

Meier, 1958) were generated using *survminer* R package to demonstrate the overall survival of risk groups stratified based on gene signature. ROC curve analysis was also performed for risk scores calculated based on selected gene signature by using *survivalROC* R package.

Significant differences in tumor stages, mutation counts and expression levels of patients in the high-risk and low-risk groups were identified using *ggstatsplot* R package. In order to validate the prognosis risk signature, the risk scores of 442 TCGA patients with LUAD were calculated using the expression values of gene signature and their coefficient values from multivariate cox proportional regression analysis. Similarly, 442 patients (after exclusion of 55 and other patients with missing data from 510 patients) were clustered into high-risk and low-risk groups and the overall survival difference between the two groups of patients was assessed by KM survival curve. Significance level used for identification of genes containing copy number variations and differentially expressed genes, was 0.01 for FDR corrected q-value. Significance level was 0.05 for FDR corrected p values (q value) for identification of genes containing significant single nucleotide variations; and was 0.05 for p-values for active subnetwork and pathway analysis, and for all statistical analysis.

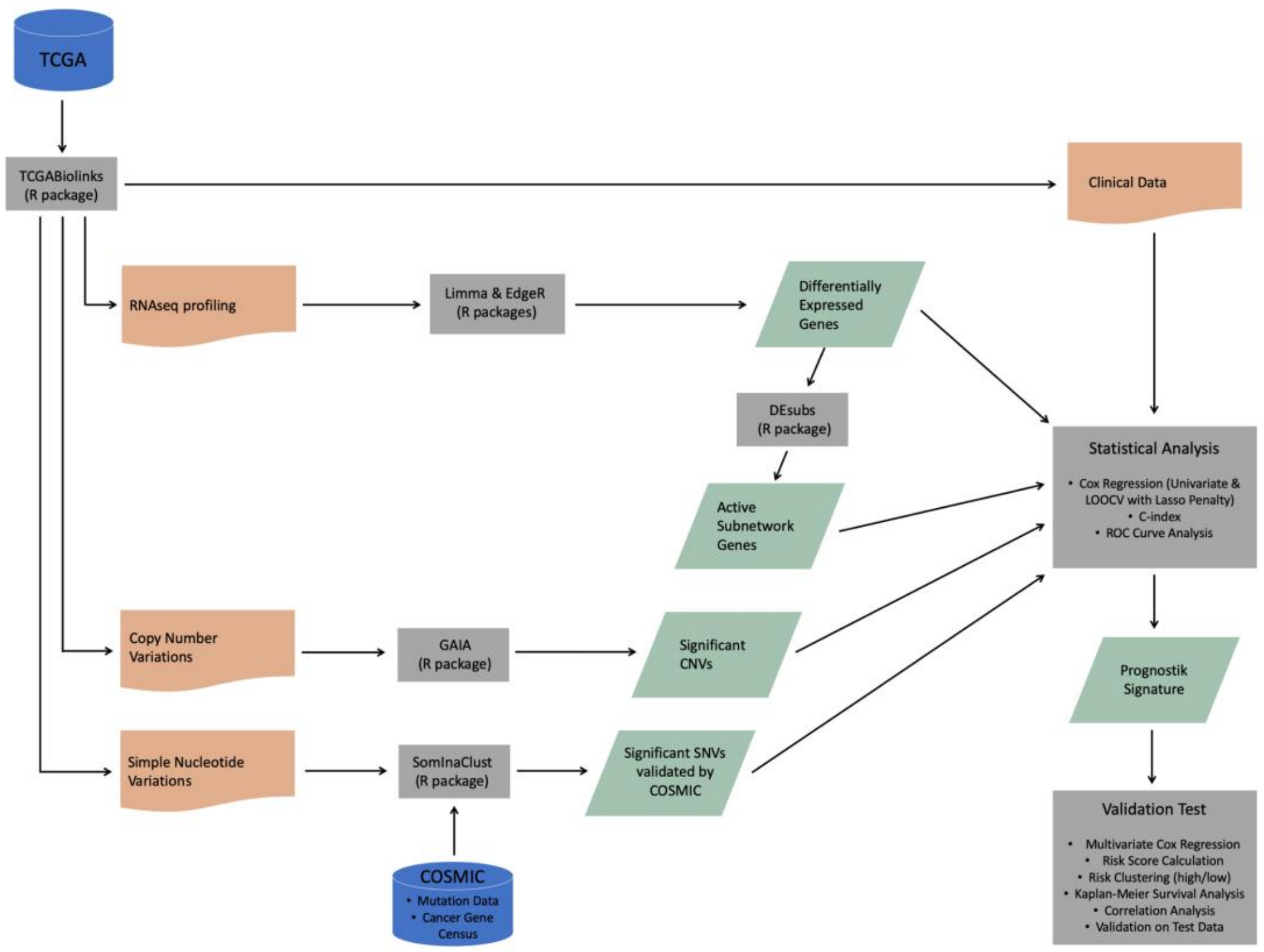

**Figure 19. The workflow of construction and validation of the prognosis gene signature**

## Abbreviations

## Declarations

### Ethics approval and consent to participate

### Consent for publication

### Availability of data and materials

The datasets supporting the conclusions of this article and the codes are available in the bitbucket repository, at https://...

## Competing interests

## Funding

## Authors' contributions

## Acknowledgements

Our grateful thanks are extended to TUBITAK ULAKBIM, High Performance and Grid Computing Center (TRUBA Resources) for the numerical calculations reported in this work.

## Authors' information

Talip Zengin

Department of Bioinformatics, Graduate School of Natural and Applied Sciences, Mugla Sitki Kocman University, Turkey

Department of Molecular Biology and Genetics, Faculty of Science, Mugla Sitki Kocman University, Turkey

talipzengin@mu.edu.tr

Tugba Önal-Süzek✉

Department of Bioinformatics, Graduate School of Natural and Applied Sciences, Mugla Sitki Kocman University, Turkey

Department of Computer Engineering, Faculty of Engineering, Mugla Sitki Kocman University, Turkey

tugbasuzek@mu.edu.tr

✉Corresponding author.